\documentclass[%
 reprint,
 amsmath,amssymb,
 aps,
floatfix,
]{revtex4-1}
\usepackage{color,soul}
\usepackage{graphicx}% Include figure files
\usepackage{dcolumn}% Align table columns on decimal point
\usepackage{bm}% bold math
\begin{document}
\newcommand{\cSe}{La$_{2}$O$_{2}$Fe$_{2}$OSe$_{2}$}
\newcommand{\cS}{La$_{2}$O$_{2}$Fe$_{2}$OS$_{2}$}
\newcommand{\cSSe}{La$_{2}$O$_{2}$Fe$_{2}$O(S, Se)$_{2}$}
\newcommand{\cM}{La$_{2}$O$_{2}$Fe$_{2}$O$M_{2}$}
\preprint{}

\title{Local Structure of Mott Insulating Iron Oxychalcogenides La$_{2}$O$_{2}$Fe$_{2}$O\textit{M}$_{2}$ (\textit{M}= S, Se)}% Force line breaks with \\%\thanks{bkfreelon@uh.edu}
%\thanks{byron.freelon@louisville.edu}%

\author{B. Karki,$^{1}$ A. Alfailakawi,$^{1}$ Benjamin A. Frandsen,$^{2,3}$  M. S. Everett,$^{4}$ J. C. Neuefeind,$^{4}$ Binjie Xu,$^{5}$ Hangdong Wang,$^{5}$ Minghu Fang,$^{5,6}$ and B. Freelon$^{1,7}$}\thanks{bkfreelon@uh.edu}
\thanks{Notice: This manuscript has been authored by UT-Battelle, LLC, under contract DE-AC05-00OR22725 with the US Department of Energy (DOE). The US government retains and the publisher, by accepting the article for publication, acknowledges that the US government retains a nonexclusive, paid-up, irrevocable, worldwide license to publish or reproduce the published form of this manuscript, or allow others to do so, for US government purposes. DOE will provide public access to these results of federally sponsored research in accordance with the DOE Public Access Plan (http://energy.gov/downloads/doe-public-access-plan).}

\affiliation{\makebox[\textwidth]{$^1$Department of Physics, University of Louisville, Louisville, Kentucky 40208, USA}\\
\makebox[\textwidth]{$^2$Materials Science Division, Lawrence Berkeley National Laboratory, Berkeley, California 94720, USA}\\
\makebox[\textwidth]{$^3$Department of Physics and Astronomy, Brigham Young University, Provo, Utah 84602, USA}\\
\makebox[\textwidth]{$^4$Spallation Neutron Source, Oak Ridge National Laboratory, Oak Ridge, Tennessee 37831, United States\\}
\makebox[\textwidth]{$^5$Department of Physics, Zhejiang University, Hangzhou 310027, China\\}
\makebox[\textwidth]{$^6$Collaborative Innovation Center of Advanced Microstructures, Nanjing University, Nanjing 210093, China}\\
\makebox[\textwidth]{$^7$Department of Physics and Texas Center for Superconductivity, University of Houston, Texas 77204, USA}\\}

%  \email{Second.Author@institution.edu}
% \affiliation{%
%  Authors' institution and/or address\\
%  This line break forced with \textbackslash\textbackslash
% }%


%  \homepage{http://www.Second.institution.edu/~Charlie.Author}
% \affiliation{
%  Second institution and/or address\\
%  This line break forced% with \\
% }%
% \affiliation{
%  Third institution, the second for Charlie Author
% \affiliation{%
%  Authors' institution and/or address\\
%  This line break forced with \textbackslash\textbackslash
% }%


\date{\today}% It is always \today, today,

\begin{abstract}
      We describe the local structural properties of the iron oxychalcogenides, La$_2$O$_2$Fe$_2$O$M_2$ (\textit{M} = S, Se), by using pair distribution function (PDF) analysis applied to total scattering data.
      %and Rietveld refinement methods applied to neutron diffraction data. 
      Our results of neutron powder diffraction show that \textit{M} = S and Se possess similar nuclear structure at low and room temperatures. The local crystal structures were studied by investigating deviations in atomic positions and the extent of the formation of orthorhombicity. Analysis of the total scattering data suggests that buckling of the Fe$_2$O plane occurs below 100 K. The buckling may occur concomitantly with a change in octahedral height. Furthermore, within a typical range of 1-2 nm, we observed short-range orthorhombic-like structure suggestive of nematic fluctuations in both of these materials. 
      %This finding points to possible ubiquity of nematic fluctuations in iron-based superconductors and related materials. DELETE THIS LAST SENTENCE}
\end{abstract}

\maketitle

%\tableofcontents

\section{\label{sec:level1}Introduction}
Iron-based superconductors (FeSCs) has attracted a large amount of attention due to the high transition temperature $T_{c}$ at which they become superconducting. Although the precise pairing mechanism in these materials remains unknown, it is thought that electron-electron interactions play an important role.
%In these materials, superconductivity (SC) is generated through an electron-electron pairing mechanism based on the  Coulomb interaction.
Such unconventional superconductivity is in contrast to the conventional electron-phonon coupling~\cite{Si2016, Paglione2010} present in BCS superconductors. In order to examine the strong-Coulombic postulate for  Fe pnictides and chalcogenides, studies were conducted~\cite{zhao2019structural,PhysRevLett.102.037003,PhysRevLett.101.076401,PhysRevB.82.024513,lee2019temperature,wong2019antiferromagnetic} to find iron-based Mott insulators that could be driven into the superconducting phase~\cite{abrahams2011quantum,shrivastava2019high}. The iron oxychalcogenides {\cM} ($M$ = S, Se) emerged as a candidate material because they are Mott insulators with structural similarities to the iron pnictides. Several attempts to induce SC in {\cSSe} have been made; however, to date, there are no published reports of SC in these systems~\cite{landsgesell2014unexpected}. Nevertheless, studying non-superconducting Mott insulators such as {\cM} can enhance our understanding of the Mott insulating region of the iron-chalcogenide electronic phase diagram~\cite{Si2016}.

Studying the Mott insulating phase in iron oxychalcogenides might enhance our understanding of the strongly correlated scenarios that lead to high temperature superconductivity \cite{PhysRevLett.104.216405}. Superconductivity in the cuprate superconductors is based on an electron or hole doping of a strongly correlated, Mott insulating phase~\cite{RevModPhys.78.17}. Hole-doped cuprates are prone to a variety of different type of electronic ordering like charge ordering and nematic ordering \cite{Fischer_2014}. The electronic nematicity breaks rotational symmetry while preserving the translational symmetry; this phase has been observed in the iron-based superconductors.
%To date, several iron pnictides and cuprates show the presence of electronic nematic states, and it was thought that the nematic phase is closely related to the superconducting mechanism in iron-based materials.
 
\begin{figure}[ht]
\centering
\includegraphics[width=0.48\textwidth]{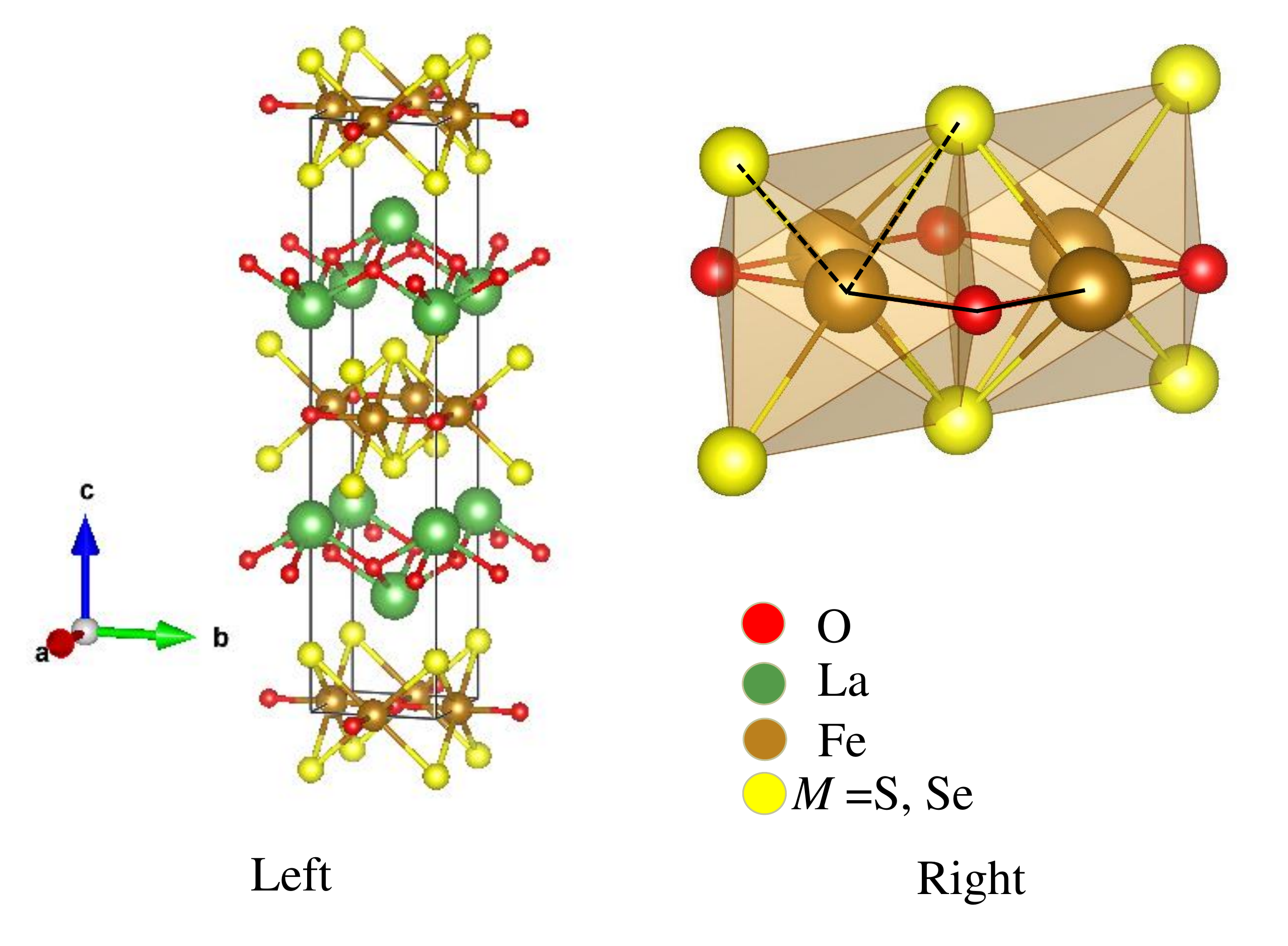}
\caption{The crystal structure of {\cM} ($M$ = S, Se) is shown in the left panel. The right panel shows FeO$_{2}$Se$_4$ octahedra, where the Fe atom is surrounded by four $M$ = S, Se  atoms and two O(2) atoms. Black solid lines show the angle formed by Fe-O-Fe atoms while the dashed lines show \textit{M}-Fe-\textit{M} angle. }
\label{fig:crystal structure}
\end{figure}

%In this paper, we study the local structure of the Mott insulating iron oxychalcogenides {\cM} where $M$ = S or Se.
The iron oxychalcogenides {\cM} were first reported as antiferromagnetic (AFM) by Mayer et. al.~\cite{doi:10.1002/anie.199216451}. This layered mixed anion material consists of a body-centered tetragonal crystal structure ($I$4/$mmm$) with fluorite-like [La$_2$O$_2$]$^{2+}$ layers and [Fe$_2$O]$^{2+}$ sheets separated by $M$$^{2-}$ anions \cite{PhysRevB.99.024109}.
% Oxygen atoms positions are non-equivalent in different layers where O(1) is for the oxygen atom in the [La$_2$O$_2$]$^{2+}$ layer and O(2) is the oxygen atom located in the [Fe$_2$O]$^{2+}$ layer.
In this structure, [Fe$_2$O]$^{2+}$  consists of an anti-CuO$_2$ arrangement with an  Fe$^{2+}$ cation coordinated by four M$^{2-}$ (above and below the plane) and two in-plane oxygen atoms, forming a tilted Fe-centered FeO$_2$Se$_4$ octahedron~\cite{doi:10.1002/anie.199216451, PhysRevB.89.100402,oogarah2018magnetic,PhysRevB.81.214433, PhysRevLett.104.216405, PhysRevB.98.045130}. Fig. \ref{fig:crystal structure} shows the crystal structure of {\cM} and its octahedra.
While an antiferromagnetic ordering was observed at N\'{e}el temperatures $T_N$ of 107.2 K and 90.1 K for 
{\cS} and {\cSe}, respectively \cite{PhysRevB.99.024109}, structural studies using X-ray \cite{PhysRevB.81.214433} and neutron \cite{PhysRevB.99.024109} powder diffraction studies did not observe a structural phase transition in {\cSSe}. Inelastic neutron scattering (INS) experiments have indicated the AFM order to be consistent with a 2 - $k$ magnetic structure in which two spin stripe phases are oriented 90$^{\circ}$ with respect to each other.~\cite{PhysRevB.99.024109, Stock_2016, PhysRevB.90.184408, PhysRevB.89.100402}
%Neutron powder diffraction studies on {\cSSe}  by Freelon et. al. \cite{PhysRevB.99.024109} shows the lack of structural phase transition on either of these materials. However, in both of these materials, they observed the antiferromagnetic ordering at a N\'{e}el temperature $T_N$ 107.2 K and 90.1 K for $M$ = S and Se, respectively. 

Therefore, the structural and magnetic behavior do not have a similar correspondence that is observed in some iron-pnictides and iron-chalcogenides in which the magnetic and structural phase transitions are in very close proximity.
%In this paper, we study the local structure of the Mott insulating iron oxychalcogenides {\cM} where $M$ = S, Se. The Mott insulating iron oxychalcogenides {\cM} were first reported as antiferromagnetic (AFM) insulator by Mayer et. al.\cite{doi:10.1002/anie.199216451} This layered mixed anion materials consists of a body-centered tetragonal crystal structure ($I$4/$mmm$) with fluorite like [La$_2$O$_2$]$^{2+}$ layers and [Fe$_2$O]$^{2+}$ sheets separated by M$^{2-}$ anions \cite{PhysRevB.99.024109}. In this structure, [Fe$_2$O]$^{2+}$  consists of anti-CuO$_2$ arrangement with  Fe$^{2+}$ cation is  coordinated by four M$^{2-}$ (above and below the plane) and two in plane oxygen atoms, forming a tilted Fe-centered FeO$_2$Se$_4$ octahedra \cite{doi:10.1002/anie.199216451, PhysRevB.89.100402,oogarah2018magnetic,PhysRevB.81.214433, PhysRevLett.104.216405}. Fig. \ref{fig:crystal structure} shows the crystal structure of {\cM} and its octahedra. The magnetic structure was studied by Freelon et. al. \cite{PhysRevB.99.024109} and the N\'{e}el temperature of {\cM} was  determined to be 107.2 K and 90.1 K respectively for $M$ = S and Se. 
The absence of a structural phase transition in the {\cSSe} Mott insulators motivates us to study the local structure to determine whether their short-range lattice symmetries undergo changes at the magnetic transitions. In particular, we are interested in understanding whether local deviations from the average tetragonal structure occur and we seek to characterize lattice correlations with physical properties.

To date, several studies of FeSCs~\cite{dai2012magnetism,Fernandes2014,chu2012divergent,li2017nematic,fernandes2013nematicity,fradkin2010nematic} have revealed the presence of a long-range, non-superconducting state called the nematic phase~\cite{Fernandes2014} that precedes the superconducting phase transition.
The long-range nematic phase, which has been observed in both iron pnictides and iron chalcogenides, develops at a structural phase transition temperature $T_s$ higher than a magnetic transition temperature $T_N$ \cite{chubukov2015origin}.
%At $T_{nem}$, which is the temperature of short-range nematic correlations,
The long-range nematic  ordering is manifested by spontaneous rotational symmetry breaking while the translational symmetry is preserved~\cite{PhysRevB.98.180505, PhysRevLett.119.187001}. 
In iron pnictide superconductors symmetry is broken between x and y directions in the Fe-plane \cite{Fernandes2014}. This reduces the rotational point group symmetry from tetragonal to orthorhombic while preserving the translational symmetry.~\cite{PhysRevB.98.180505,fradkin2010nematic}.
A hallmark feature of the nematic phase in the FeSCs is the presence of large anisotropies of electronic properties, such as the resistivity, along the x and y directions. These anisotropies are larger than would be expected from the magnitude of the structural distortion alone, indicating that the nematic instability exists in the electronic system, while the structural symmetry breaking is simply a consequence of the underlying electronic symmetry breaking. Similar electronic nematicity has also been observed in other electron fluids such as those observed in the cuprates or quantum Hall systems \cite{PhysRevB.82.180511,fradkin2010nematic}. This electronic nematicity is related to, but distinct from, other types of nematicity, such as the original use of the term to describe the directional alignment of molecules in liquid crystals \cite{PhysRevB.82.180511}, or a purely structural nematic order that is unrelated to any instabilities of an electron fluid.

In the case of FeSCs, the nematic state is suppressed upon doping, and superconductivity eventually emerges out of the nematic phase \cite{kushnirenko2018superconductivity, koch2019room}. Sometimes, nematic ordering persists into the superconducting state, as in LiFeAs \cite{Yim2018}. The origin of the electronic nematic ordering is not yet fully understood. In the pnictides, it appears to be related to the magnetism, but this may not be the case in iron-chalcogenides such as FeSe, which does not show long-range magnetic ordering \cite{chubukov2015origin}.
Recently, the pair distribution function (PDF) technique~\cite{Egami2003,Farrow_2007} has been used to study the local atomic arrangement and  structural fluctuations frequently associated with nematic fluctuations in iron-based superconductors for both normal and superconducting phases. PDF methods can be used to measure local scale deviations from the global crystal symmetry of a material~\cite{frandsen2019quantitative, Fernandes2014}.
%Recent efforts to study structural nematic fluctuations have used the pair distribution function (PDF) technique~\cite{Egami2003,Farrow_2007} to measure the local structure fluctuation from the average structure \cite{frandsen2019quantitative, Fernandes2014}.
In FeSe$_{1-x}$Te$_{x}$ systems, PDF investigated the reduction of local crystal symmetry which enhances the local moment formation leading to magnetic instability \cite{PhysRevB.81.134524}. X-ray and neutron PDF on FeSe and (Sr, Na)Fe$_{2}$As$_{2}$ revealed the presence of short-range orthorhombic distortions at temperatures well above the static nematic and orthorhombic ordering temperature $T_s$~\cite{koch2019room, frandsen2019quantitative, PhysRevLett.119.187001, PhysRevB.98.180505}. These short-range orthorhombic fluctuations of the structure were interpreted as consequences of fluctuations of the underlying electronic nematic phase.
%X-ray and  PDF on the FeSe superconductor confirmed the $C_4$ symmetry breaking in the 122 family of iron pnictide superconductor and the structural fluctuations \cite{koch2019room, frandsen2019quantitative}.
The observation of structural nematic fluctuations in iron-based superconductors, by using neutron and x-ray PDF analysis, provided evidence that nematic degrees of freedom exist near the superconducting phase \cite{PhysRevB.98.180505}. PDF analysis has also been used by Horigane et. al~\cite{ horigane2014local} to probe the local structure of {\cS}  and found a large anisotropic thermal displacement parameter for the O(2) atom along the $c$-axis ($U_{33}$). That study concluded that the crystal structure is strongly coupled to the magnetism in this system.

%In this paper, we study the local structure of the Mott insulating iron oxychalcogenides {\cM} where $M$ = S or Se. The iron oxychalcogenides {\cM} were first reported as antiferromagnetic (AFM) insulator by Mayer et. al.\cite{doi:10.1002/anie.199216451} This layered mixed anion materials consists of a body-centered tetragonal crystal structure ($I$4/$mmm$) with fluorite like [La$_2$O$_2$]$^{2+}$ layers and [Fe$_2$O]$^{2+}$ sheets separated by M$^{2-}$ anions \cite{PhysRevB.99.024109}. It is worth mentioning that oxygen atoms positions are non-equivalent different layers while O(1) is for the oxygen atom in the [La$_2$O$_2$]$^{2+}$ layer,  O(2) is the atom located in the [Fe$_2$O]$^{2+}$ layer. In this structure, [Fe$_2$O]$^{2+}$  consists of anti-CuO$_2$ arrangement with  Fe$^{2+}$ cation is  coordinated by four M$^{2-}$ (above and below the plane) and two in plane oxygen atoms, forming a tilted Fe-centered FeO$_2$Se$_4$ octahedra \cite{doi:10.1002/anie.199216451, PhysRevB.89.100402,oogarah2018magnetic,PhysRevB.81.214433, PhysRevLett.104.216405}. Fig. \ref{fig:crystal structure} shows the crystal structure of {\cM} and its octahedra. The magnetic structure was studied by Freelon et. al. \cite{PhysRevB.99.024109} and the N\'{e}el temperature of {\cM} was  determined to be 107.2 K and 90.1 K respectively for $M$ = S and Se.

In this work, we report the use of the neutron pair distribution function (PDF) technique to study the local structure of {\cSSe}. The experimental methods are discussed in Section II, the results of this study are presented in Section III, section IV includes the discussion and the conclusions are presented in section V. Our results indicate the presence of short-range orthorhombic distortions suggestive of short-range nematicity in both materials \cSSe, despite the persistence of tetragonal symmetry in the average structure at all temperatures. 
%Our results suggest the presence of structural nematic fluctuations in both materials {\cSSe}. 
We discuss the behavior of these fluctuations as a function of temperature. It appears that the change in the chalcogen, from S to Se, does not affect this behavior. 
%\textcolor{red}{The energy-integrated nature of the PDF measurements means we are probing fluctuations on a time scale of 10$^{-13}$ s or slower. However, we were unable to extract any other temporal information using the PDF data alone \cite{PhysRevLett.119.187001}. Complementary probes such as NMR or Moessbauer spectroscopy could help clarify the situation \cite{Freelon2021}.}

\section{\label{sec:level2}Experimental methods}

\subsubsection{Synthesis}
{\cSSe} samples were prepared by a solid-state reaction method \cite{doi:10.1002/anie.199216451} from stoichiometric amounts of high purity La$_2$O$_3$, S, Se, and Fe powders. The appropriate mixture of these powders was grounded thoroughly, pelletized then heated in an evacuated quartz tube at 1030$^{\circ}$ C for 72 hours. This process was repeated three times. After being pulverized into fine powders, lab-based x-ray powder diffraction showed the materials to be of the high-quality and single phase \cite{doi:10.1002/anie.199216451, PhysRevB.99.024109,PhysRevLett.104.216405}.

\subsubsection{Neutron Powder Diffraction}
Time-of-flight (TOF) neutron powder diffraction measurements were performed at the POWGEN diffractometer of the Spallation Neutron Source (SNS) at Oak Ridge National Laboratory (ORNL). Five grams of {\cS} and 4.5 g of {\cSe} powders were placed in vanadium cans. Room temperature (300 K) measurement were performed, on each sample, for approximately one hour.
%Seven measurements at the temperatures 10, 50, 100, 150, 200, 250 and 300 K were performed, on each sample, for approximately one hour. 
The center wavelength and $d$-spacing range of neutrons in this experiment were 1.5 \AA\ and 0.49664-13.50203 \AA\, respectively. Rietveld refinements were conducted using GSAS-II software~\cite{Toby:aj5212}.

\subsubsection{Pair Distribution Function}
Neutron total scattering data were collected using the Nano-scale Ordered Materials Diffractometer (NOMAD) beamline at the SNS in ORNL. {\cSSe} powders were loaded and sealed in the vanadium cans inside a glove-box and were placed in an orange cryostat. A total of sixteen measurements were performed on each sample at various temperatures ranging from 2 K to 300 K. A higher number of data points were taken, in small steps, around the N\'{e}el temperature. Data were collected for approximately two hours at each temperature. In order to obtain structural information relevant to the samples, an empty vanadium sample can was measured for two hours for background subtraction.

%By Fourier transforming the total scattering data, the atomic pair distribution function (PDF)  was obtained~\cite{Egami2003} and was calculated by
%\begin{equation} 
%g(r)=1+\frac{1}{2\pi^2 r \rho}\int_{Q_{min}}^{Q{max}}Q[S(Q)-1]sin(Qr)\ dQ
%\end{equation}
%where $g(r)$ is the pair distribution function, $\rho(r)$ is the microscopic pair density
The reduced pair distribution function (PDF) $G(r)$ is obtained experimentally by Fourier transforming the total scattering data as follows:
\begin{equation}
 G(r) = \frac{2}{\pi} \int_{Q_{min}}^{Q{max}}Q(S(Q) - 1)sin(Qr)dQ,  
\end{equation}

Where $S(Q)$ is the normalized structure factor, and $Q$ is the scattering vector, defined as $Q=\frac{4 \pi sin(\theta)}{\lambda}$ in which $\lambda$ and $\theta$ are the neutron wavelength and scattering angle, respectively. PDF data were refined and analyzed using \texttt{PDFGUI}~\cite{Farrow_2007}  program and \texttt{DIFFPY-CMI}~\cite{Juhas:po5132} suite. 
For the PDF analysis, values for $Q_{min}$ and $Q_{max}$ were 0.1 \r{A} and 31.4 \r{A} respectively. %The reduced pair distribution function $G(r)$ is related to the atomic pair distribution function by equation~(\ref{equation:G_r}), where $\rho_0$ is the average number density. 
The calculated reduced pair distribution function $G_{c}(r)$ is obtained from the structural model using equation~(\ref{equation:G_cal})~\cite{Proffen:gl0603}.

%\begin{equation}
 %   G(r)=4\pi r \rho_0\ (g(r)-1)
%\label{equation:G_r}
%\end{equation}

\begin{equation}
G_{c}(r)= \frac{1}{r}\sum_{i}\sum_{j} \Big[\Big(\frac{b_ib_j}{<b>^2}\Big)\delta(r-r_{ij}) \Big]- 4\pi r\rho_0
\label{equation:G_cal}
\end{equation}
where $b_i$, $b_j$ and $<b>$ are the average scattering power of constituent atoms of the sample. The summation is over all atoms $i$ and $j$ in the model, where the distance separating any given pair of atoms is $r_{ij}$.

\section{\label{sec:level3}Results}
\subsection{\label{sec:Crystal structure}Average and Local Structure}
\begin{figure}
\includegraphics[width=3.4in]{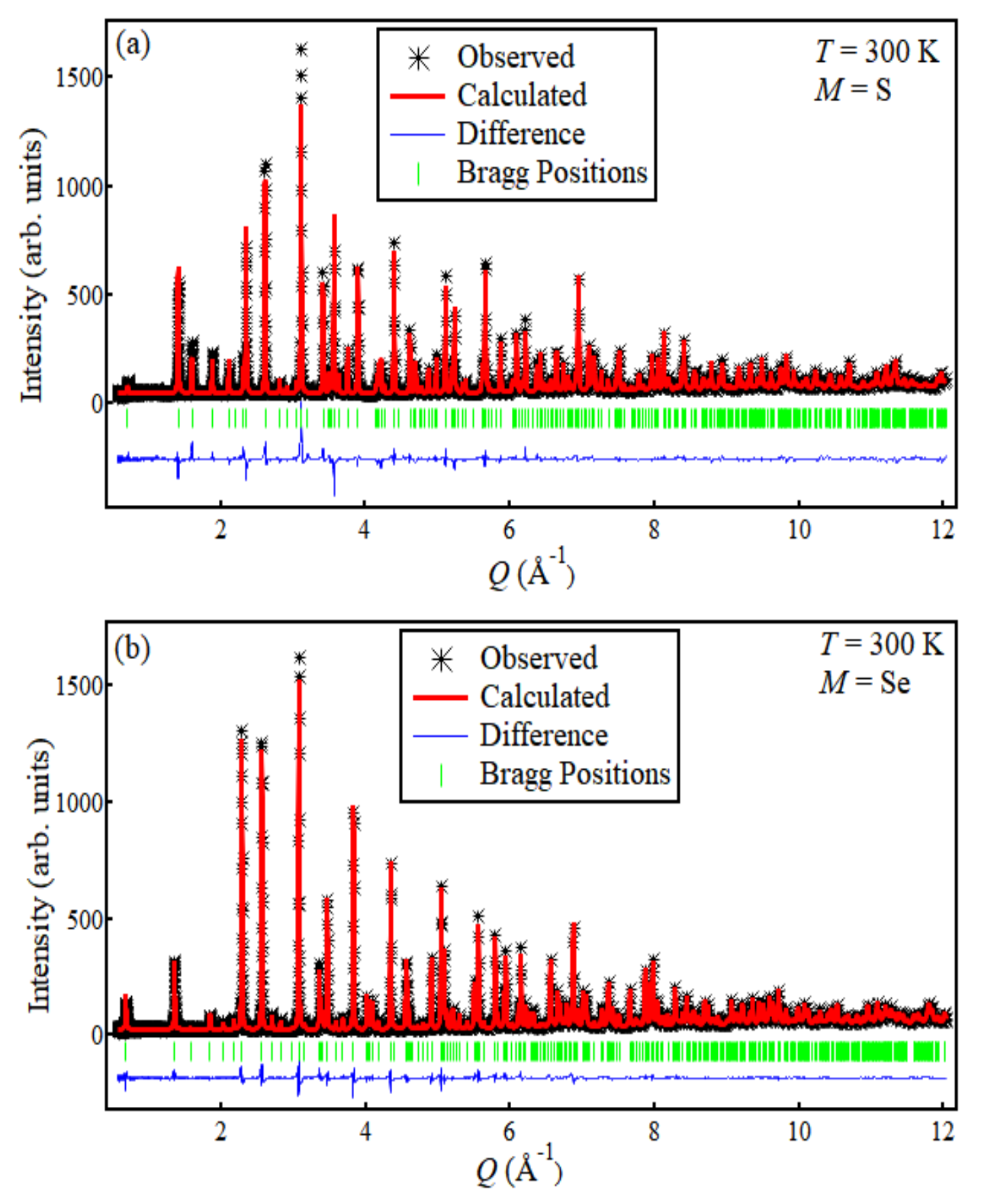}
\caption{Neutron Rietveld refinement of the nuclear model of a) $M$ = S and b) $M$ = Se at 300 K as a function of momentum transfer $Q$. The details of the fit are provided in the text.}
\label{Rietveld refinement}
\end{figure}

%\subsection{\label{sec:Crystal structure}Crystal Structure and Lattice Parameters}
We confirmed the crystal structure of our samples through the neutron powder diffraction. The Rietveld refinement of neutron powder diffraction data was performed using GSAS II software \cite{Toby:aj5212}.  Fig. \ref{Rietveld refinement}(a) and (b) show the Rietveld refinement of neutron powder diffraction data measured at 300 K using neutrons of central wavelength 1.5 \r{A}. Both $M$ = S, Se materials show similar nuclear structures having space group $I4/mmm$ (space group No. 139) which is consistent with the previous studies \cite{doi:10.1002/anie.199216451,PhysRevB.99.024109}. Our findings show that the lattice parameters of {\cSe} ($a$ = 4.0887(5) \r{A}, $c$ = 18.6081(3) \r{A}) is larger than that of the {\cS} ($a$ = 4.0439(9) \r{A}, $c$ = 17.8945(6) \r{A}), as expected for the larger Se atoms. Rietveld refinement parameters of both $M$ = S, Se are tabulated in Table \ref{tab:RRparameter} including lattice parameters ($a$ and $c$) as well as the anisotropic thermal displacement parameters $U_{33}$  for all atoms. %\textit{Karki, recall the reviewer's comments when we did not directly reference the tabulated parameters.  It is not at all necessary --as known from lots of published papers-- but please describe the tabulated parameters in the paragraph e.g., Rs, Bs etc...} 
It was observed that $U_{33}$ for O(2) site is much larger than all the other sites. 
For completeness Table \ref{tab:atompositions} presents the atomic positions of constituents La$_{2}$O$_{2}$Fe$_{2}$O$M_{2}$ ($M$ = S, Se) with their respective Wyckoff symbols and occupancies.
The quality of the presented fits was characterized by the listed Rietveld parameters: $R$-whole pattern $R_{w}$, crystallographic factor $R_{F}$ and goodness of fit tabulated in Table \ref{tab:RRparameter}.

\begin{table}[h!] \centering
\begin{tabular}{ccccccc}
\hline
\hline
\text{}&\text{\cSe} &&\text{\cS}   \\
\hline 
{$a$} (\AA)& 4.0887(5)    &&4.0439(9) \\
{$c$} (\AA)& 18.6081(3) &&17.8945(6)\\
{La - $U_{11}$(\AA$^2$)}&   0.0028(8) && 0.00123(9)\\
{Fe - $U_{11}$}(\AA$^2$)&   0.0065(4) && 0.00729(5)\\
{O(1) - $U_{11}$}(\AA$^2$)&   0.0038(4) && 0.00013(4)\\
{O(2) - $U_{11}$}(\AA$^2$)&   0.0037(7) && 0.00510(6)\\
{$M-U_{11}$}(\AA$^2$)&   0.0066(3) && 0.00319(8)\\
{La - $U_{33}$}(\AA$^2$)&   0.00722(5) && 0.01718(6)\\
{Fe - $U_{33}$}(\AA$^2$)&    0.01077(5) && 0.02258(8)\\
{O(1) - $U_{33}$}(\AA$^2$)&    0.00998(8) &&0.02049(9)\\
{O(2) - $U_{33}$}(\AA$^2$)&    0.02226(9) &&0.04643(5)\\
{$M-U_{33}$}(\AA$^2$)&    0.00522(5) && 0.02124(9)\\
{Volume} (\AA$^3$)&311.08(8) &&292.64(1)\\
{Space Group}&$I_{4}/mmm$&&$I_{4}/mmm$ \\
$R_{w}$(\%)&6.74&&6.79\\
$R_{F}$(\%)&2.71&&6.57\\
Goodness of fit &4.73&&6.57\\
\hline\hline
\end{tabular}
\caption{Refined parameters from Rietveld analysis of neutron powder diffraction data of {\cSSe} at $T$ = 300 K.} %$B$ is the isotropic thermal parameter}
\label{tab:RRparameter}
\end{table}

\begin{table*}
%\footnotesize\baselineskip=8pt
\begin{center}
\begin{tabular}{lccccccccccccc}
\hline 
\hline
\text{} &&& La$_{2}$O$_{2}$Fe$_{2}$OSe$_{2}$	&&&&&& La$_{2}$O$_{2}$Fe$_{2}$OS$_{2}$\\
\hline
	%$^{\mbox{b}}$
   Atom & Site & Occupancy & 
  x   &   y  &  z  &  Atom    &  Site & Occupancy  &  x    &  y     &  z &  \\
\hline
\textbf{La}                               & 4e & 0.97(5) & 
  0.5000  &  0.5000 &   0.1845(6) & \textbf{La}  & 4e & 0.94(7) &  0.5000 &  0.5000 & 0.1806(7) \\
\textbf{Fe}                       & 4c & 0.94(5) & 
  0.5000 &  0.0000 &  0.0000 &   \textbf{Fe} & 4c & 0.92(7) &   0.5000 &  0.0000   &  0.0000 \\
\textbf{Se}                       & 4e & 0.95(6) &
  0.0000 &  0.0000 &  0.0963(6) & \textbf{S} & 4c & 0.92(8) & 0.0000 &  0.0000   &  0.0945(13) \\
\textbf{O1}                       & 4d & 0.98(6) & 
  0.5000 &  0.0000 &  0.2500 &   \textbf{O1} & 4d & 0.94(9) &   0.5000 &  0.0000   &  0.2500 \\
\textbf{O2}                      & 2b & 0.97(7) &
  0.5000 &  0.5000 &  0.0000 &   \textbf{O2} & 2b & 0.96(6) &   0.5000 &  0.5000   &  0.0000 \\

\hline
\hline
\end{tabular}
\end{center}
\caption{\label{tab:atompositions}Atomic site, occupancy and fractional atomic coordinates of La$_{2}$O$_{2}$Fe$_{2}$O(S, Se)$_{2}$  at $T$ = 300 K extracted from Rietveld analysis.}
\end{table*}

%PDF fit quality
\begin{table*}
%\footnotesize\baselineskip=8pt
\begin{center}
\begin{tabular}{ccc|ccccccc}
\hline 
\hline
\text{} &La$_{2}$O$_{2}$Fe$_{2}$OSe$_{2}$	&&& La$_{2}$O$_{2}$Fe$_{2}$OS$_{2}$\\
\hline
	%$^{\mbox{b}}$
    & Pseudo-Orthorhombic & Tetragonal &   &  Pseudo-Orthorhombic   &  Tetragonal &  \\
\hline
\textbf{$\chi^2$}    & 24.20 & 43.77 && 
  41.45  &  74.13 \\
\textbf{$Reduced-\chi^2$}    & 0.03 & 0.05 && 
  0.05  &  0.09 \\
\textbf{$R_w$}    & 0.11 & 0.16 && 
  0.14  &  0.20 \\
\hline
\hline
\end{tabular}
\end{center}
\caption{\label{tab:PDFfitquality}Comparison of the PDF fit quality parameters of pseudo-orthorhombic and tetragonal models of room temperature data within the range 1.8 \r{A} ${\le}\ r\ {\le}$ 20.0 \r{A}.}
\end{table*}

%\begin{figure}[!ht]
%\includegraphics[width=3.0in]{Fig3-edited.pdf}
%\caption{PDF refinement of $M$ = Se fitted with pseudo-orthorhombic model at a) 300 K and b) 2 K within the range of 1.8 \r{A} ${\le}\ r\ {\le}$ 4.5 \r{A}. %At 300 K, the first, second and third peaks are mainly comprised of Fe-O(2), La-O(1) and Fe-Se correlation, respectively. However, at a lower temperature, the most noticeable difference is the clear propelling of the third peak, which is the correlation of Fe-Se, into Fe-Se at 2.67 \r{A} and O(1)-O(1) at 2.87 \r{A}. 
%Observed and calculated PDF patterns are represented by G$_{obs}$ (black) and G$_{calc}$ (red), respectively, with the difference profile by G$_{diff}$ (green).}
%\label{peak splitting}
%\end{figure}

\begin{figure}[!ht]
\includegraphics[width=3.4in]{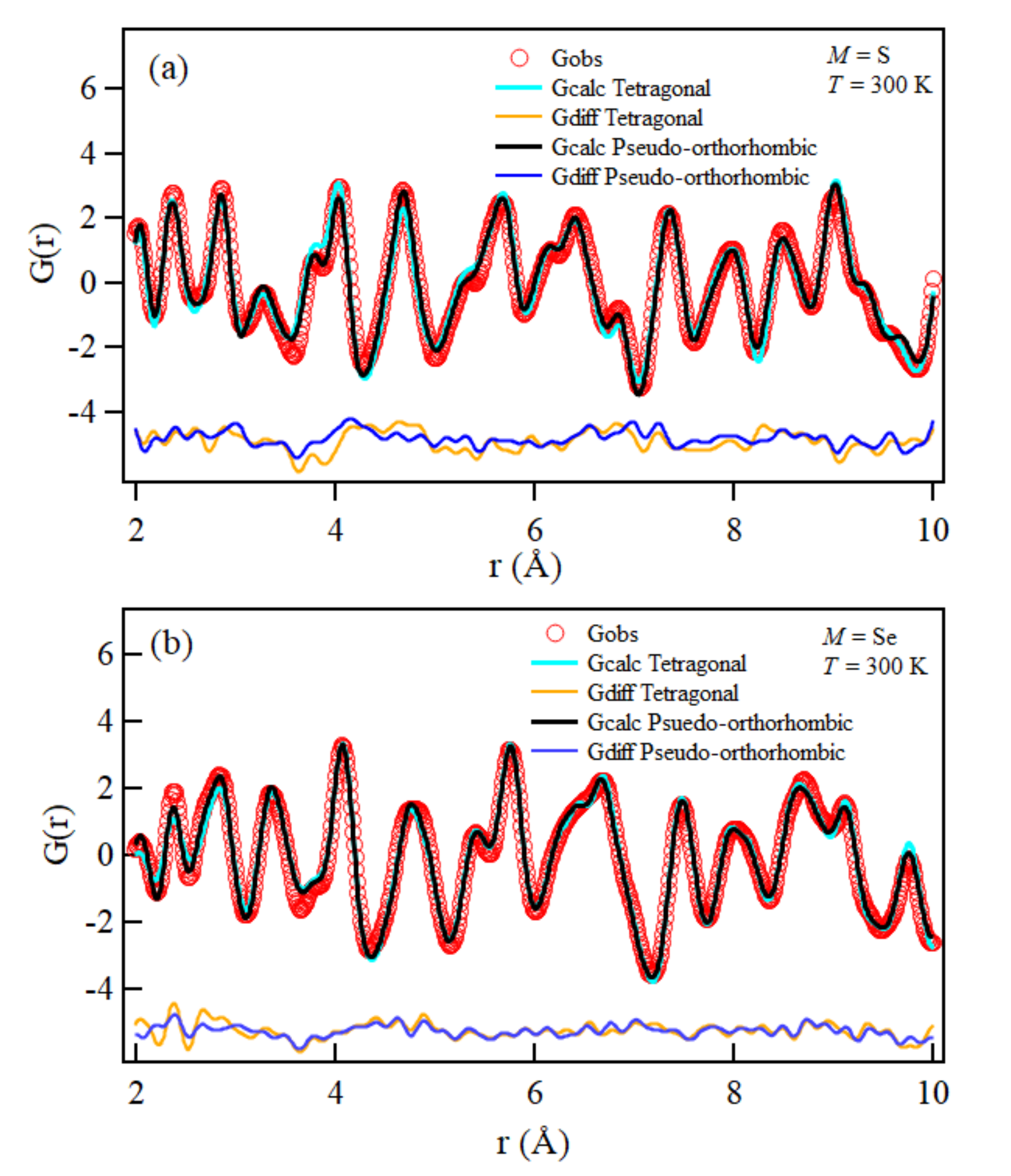}
\caption{Room temperature PDF fits of a) $M$ = S and b) $M$ = Se fitted with both tetragonal and pseudo-orthorhombic model within the range of 1.8 \r{A} ${\le}\ r\ {\le}$ 10.0 \r{A}.}
\label{com pdf}
\end{figure}

\begin{figure}[htp]
\includegraphics[width = 2.5 in]{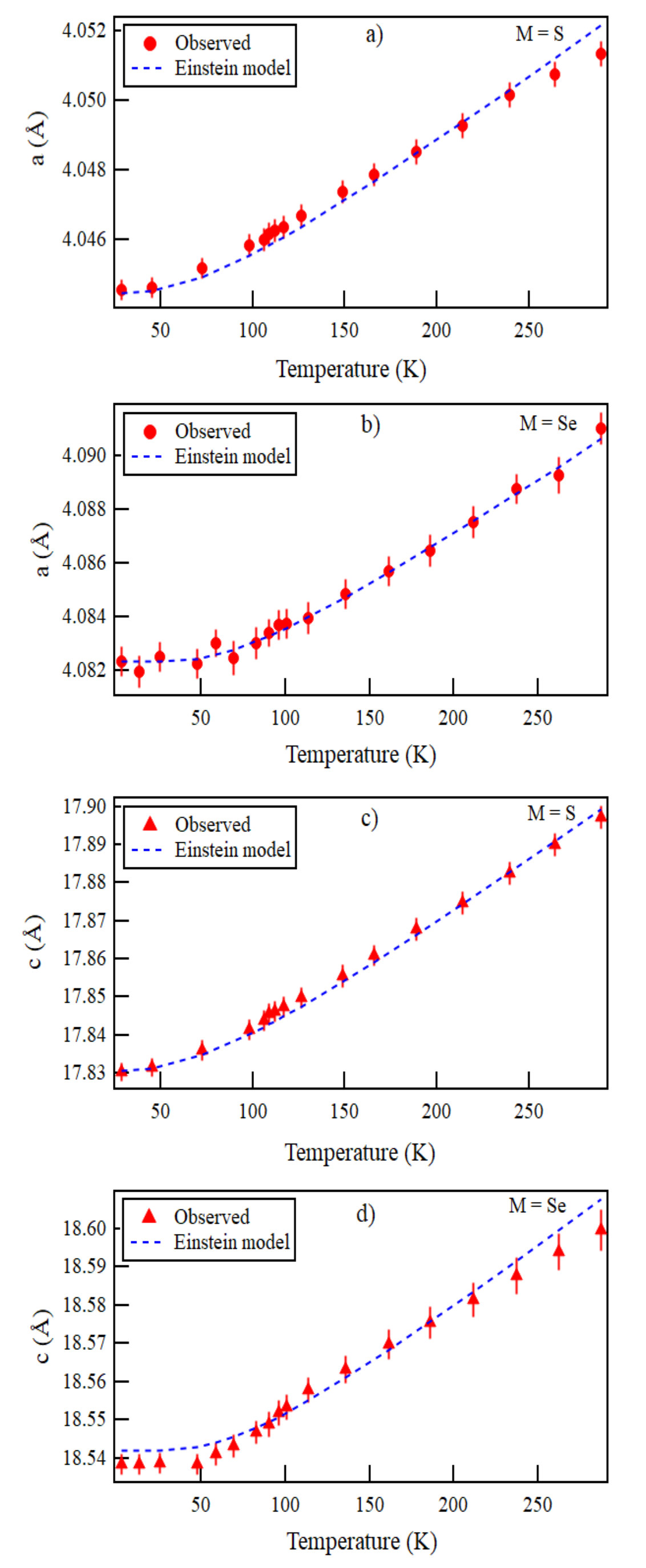}
\caption{Lattice parameters as a function of temperature for $M$ = S, Se  determined from temperature-series sequential PDF refinement. a) $a$- and c) $c$-lattice parameter are shown for $M$ = S and b) $a$- and d) $c$-lattice parameter are shown for $M$ = Se. Blue dotted lines represent an Einstein model of lattice thermal expansion (see text).}
\label{lattice parameter}
\end{figure}
Having obtained a confirmation of the global lattice symmetry for our samples, we then investigated the local symmetry~\cite{koch2019room, PhysRevB.98.180505,white2010effects,pasciak2019local,slawinski2019neutron,jeong2019high}. The local structure data of  La$_{2}$O$_{2}$Fe$_{2}$O$M_{2}$ ($M$ = S, Se) was analyzed by using the PDF refinement program \texttt{PDFGUI} ~\cite{Farrow_2007} to determine the lattice parameters, scale factor, atomic positions, and the thermal parameters. In order to allow for local orthorhombic symmetry breaking, we parameterized the lattice constants as $a = a_{mid} (1+\delta)$ and $b = a_{mid} (1-\delta)$, where $a_{mid}=\frac{a+b}{2}$ and $\delta = |\frac{a-b}{a+b}|$. The orthorhombicity parameter, $\delta$ was refined with a pseudo-orthorhombic model while in tetragonal model $\delta$ is 0 as lattice parameter $a$ and $b$ were equal. All refinements were performed on the Nyquist grid. Instrumental parameters $q_{damp}$ and $q_{broad}$, which are the parameters that correct the PDF envelope function~\cite{masadeh2007quantitative,toby1992accuracy} for the instrument resolution~\cite{Proffen:gl0603, Farrow_2007}, were fixed by using an independent measurement of a standard Si sample. 
A direct comparison of quality of fit parameters of pseudo-orthorhombic symmetry and tetragonal symmetry PDF fits of room temperature data of La$_{2}$O$_{2}$Fe$_{2}$O$M_{2}$ ($M$ = S, Se) is shown in Table \ref{tab:PDFfitquality} and Fig. \ref{com pdf}. It is clear that pseudo-orthorhombic model provides a significant improvement to the fit over the tetragonal model. This suggests that the local structure undergoes short-range orthorhombic distortions, while the average structure captured by Rietveld refinement against the Bragg peaks is purely tetragonal.

%In Fig.~\ref{peak splitting}, refined PDF data is presented for {\cSSe}.  The observed data G$_{obs}$ is shown in black and the calculated data G$_{calc}$ is shown in red, while the difference between the observed data and the calculated data, G$_{diff}$, is shown in green. We observed that the third peak of Fig.~\ref{peak splitting}a, representing a Fe-Se distance of  2.72 \r{A}  at 300 K, shows a clear modification into Fe-Se at 2.67 \r{A} and Fe-Fe, O(1)-O(1) at 2.87 \r{A} at low temperature (see Fig.~\ref{peak splitting}b). The peak change indicates two slight changes of Fe and O positions that affect the associated interatomic positions. The implications of these atomic displacements are discussed below.

The lattice constants were obtained as a function of temperature using PDF analysis. Fig.~\ref{lattice parameter} shows the temperature dependence of the lattice parameters $a$ and $c$ for both {\cSSe} samples in the range of 1.8 to 49.99 \r{A}, where the Einstein model was superimposed to such obtained $a$($T$) and $c$($T$) dependencies. Following previous work~\cite{PhysRevB.81.214433}, the Einstein fitting model was calculated using  equations~(\ref{equation:E1}) and (\ref{equation:E2}) under the assumption that the thermal expansion is proportional to the internal energy of a quantum mechanical oscillator~\cite{einsteinplot}. Fig. \ref{lattice parameter} also shows that the lattice parameter $a$ increases with increasing temperature and fits well with the Einstein model of thermal expansion~\cite{einsteinplot}. However, the lattice parameter $c$ shows a kink near 90 K in the case of $M$ = Se (see discussion). 
%However, lattice parameter $c$ show discontinuities near 90 K and 107 K for $M$ = Se and S, respectively.
\begin{equation}
a(T) = a_{0}\Big[1+\alpha\, \theta_{E} f_{E}\Big(\frac{\theta_{E}}{T}\Big) \Big]
\label{equation:E1}
\end{equation}
\begin{equation}
f_{E}\Big(\frac{\theta_{E}}{T}\Big)=\frac{1}{\exp(\frac{\theta_{E}}{T})-1}  
\label{equation:E2}
\end{equation}
Here, $a$ is the lattice constant, $a_{0}$ and $c_{0}$ are lattice constants at low temperature (2k for $M$ = Se and 30 k for $M$ = S system) obtained from PDF fits, $\alpha$ is the high temperature thermal expansion coefficient, and $\theta_{E}$ is the characteristic Einstein temperature. In our plots, we have used $\theta_{E}$ = 211 K as reported by Free \textit{et al.} \cite{PhysRevB.81.214433}.

%\begin{figure}[!h]
%\includegraphics[width=4.3in]{Fig5_12.pdf}
%\caption{Panel a) shows possible Fe and Se/S atomic movement %within the {\cSSe} octahedra. The dashed orange lines in b) %show the Fe atom movement in the Fe$_2$O plane and the solid %black lines show Fe movement with respect to $M$.}
%\label{fig:atom_move}
%\end{figure}

\subsection{\label{sec:ADPs}Temperature Dependence of Thermal Displacement Parameters}
\begin{figure}[!h]
\centering
\includegraphics[width=2.5 in]{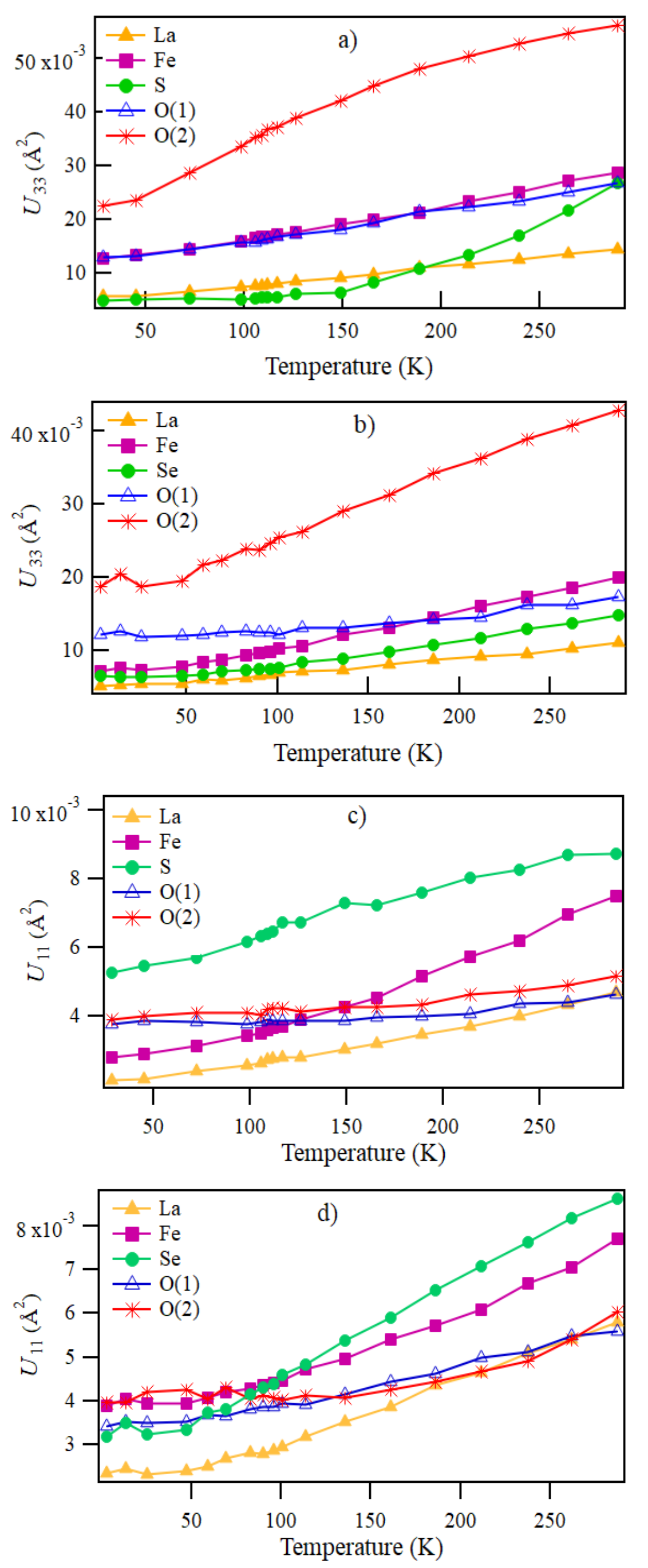}
\caption{The temperature dependence of the anisotropic thermal displacement parameters a) $U_{33}$ and c) $U_{11}$ for $M$ = S; and b) $U_{33}$ and d) $U_{11}$ for $M$ = Se for all atomic sites determined from temperature-series sequential PDF refinement within the range 1.8 \r{A} ${\le}\ r\ {\le}$ 20.0 \r{A}.}
\label{fig:ADPs}
\end{figure}

Fig.~\ref{fig:ADPs} shows the temperature dependence of the anisotropic thermal displacement parameters $U_{11}$ and $U_{33}$ for both $M$ = S, Se from PDF fits. We note the relatively high values of $U_{11}$ parameter for the $M$ = S, Se sites and the $U_{33}$ parameter for the O(2) site. The large $U_{33}$ parameter for the O(2) site agrees well with the neutron powder diffraction results (See Table \ref{tab:RRparameter}). The large an anisotropic thermal displacement parameter $U_{33}$ corresponds to O(2) displacement above and below the Fe$_2$O plane. This particular displacement can be accompanied by a distortion of Fe-O-Fe angle (See Fig.~\ref{fig:crystal structure} where this angle is highlighted by a solid black line in the octahedra). Another possible movement along the $c$-axis is that of  sulfur/selenium atoms with respect to the Fe atoms in the octahedra. This motion would affect the $M$-Fe-$M$ angle in the octahedra (See Fig.~\ref{fig:crystal structure} where this angle is highlighted by dashed black lines in the octahedra). 

\begin{figure}[!h]
\includegraphics[width=0.28\textwidth]{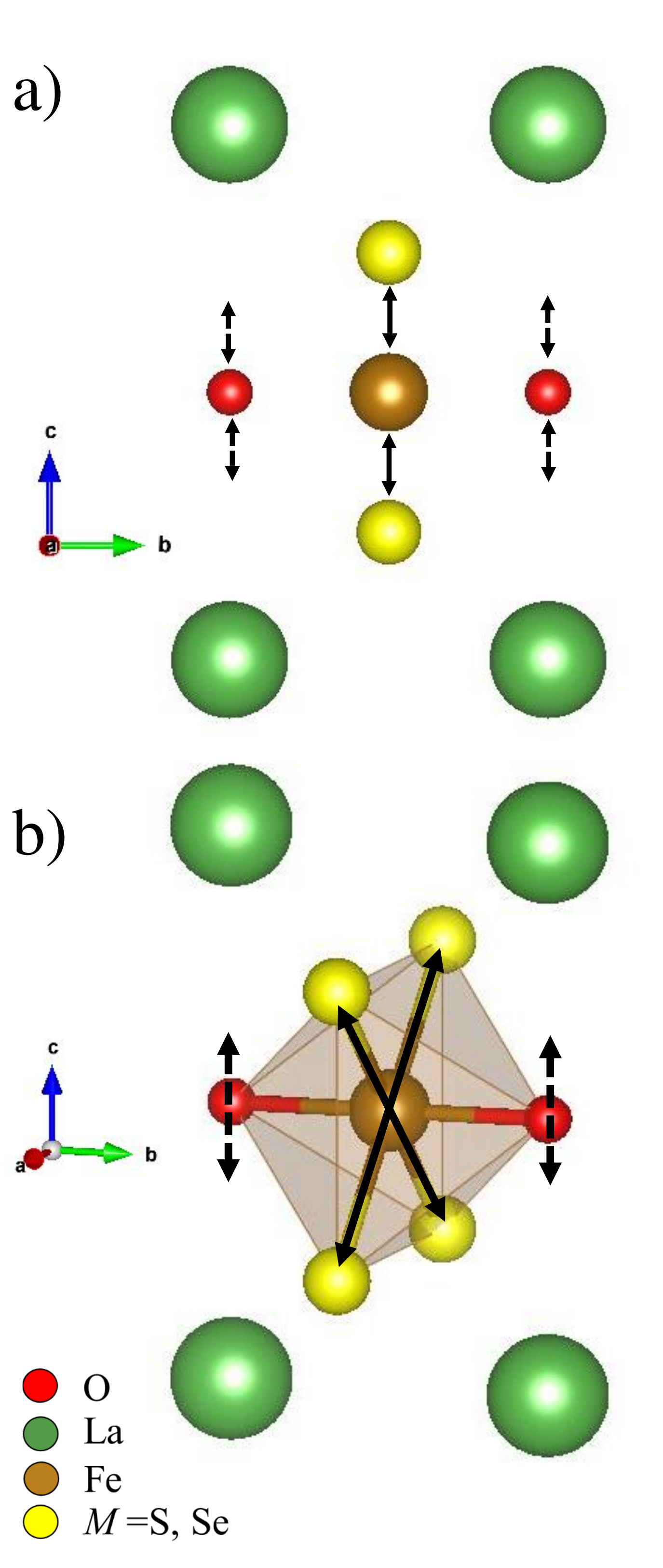}
\caption{Panel (a) shows possible Fe and Se/S atomic movement within the {\cSSe} octahedra that result in the bucking of the Fe$_2$O plane. (b) The dashed lines show Fe atom movement in the Fe$_2$O plane and the solid lines show Fe movement with respect to $M$.}
\label{fig:atom_move}
\end{figure}

Fig.~\ref{fig:atom_move} provides views of the atomic displacements under discussion.  Panel a) shows a view along the $a$-axis while panel b) provides an isometric perspective view by presenting slight rotations about the $a$ and $c$ crystal axes. 
In panel b) black solid lines show the Fe atom movement and black dashed line shows the movement of O(2) atom.
%\textcolor{red}{(I am not sure whether Alaa worked on Ben's comment on this text or not)}
%In panel b) the Fe-O-Fe angle is indicated with solid lines and the $M$-Fe-$M$ angle is represented with  dashed lines. 
When the O(2) atoms in the Fe$_2$O plane move up or down the angle  Fe-O-Fe changes causing buckling of the Fe$_2$O plane. O(2) atoms are displacing out of the plane by $\sim$ 0.015 (2) \r{A}.
Allowing the $z$-coordinates of the two O(2) atoms to displace opposite to each other results in the $U_{33}$ parameter for the O(2) site dropping down to a value that is in line with other atomic sites. This result suggests that local buckling of the Fe$_2$O plane may occur through small displacements of the O(2) atom above and below the plane.
%Therefore, the atomic displacement parameters provide insight into how buckling might occur.
%(need to includes O(2) displacement magnitudes and direction
%The modified form factor, $f_{modified}$ is the $f_{atomic}$ with correction due to temperature, and it is calculated using $f_{modified}=f_{atomic}\ e^{-\frac{B\ sin^2\theta}{\lambda^2}}$. The Debye Waller factor $B$ is related to $U$ by $B=8\ \pi^2<U^2>$ \r{A}, $U$ in general is anisotropic.

\begin{figure}[ht]
\includegraphics[width=3.0in]{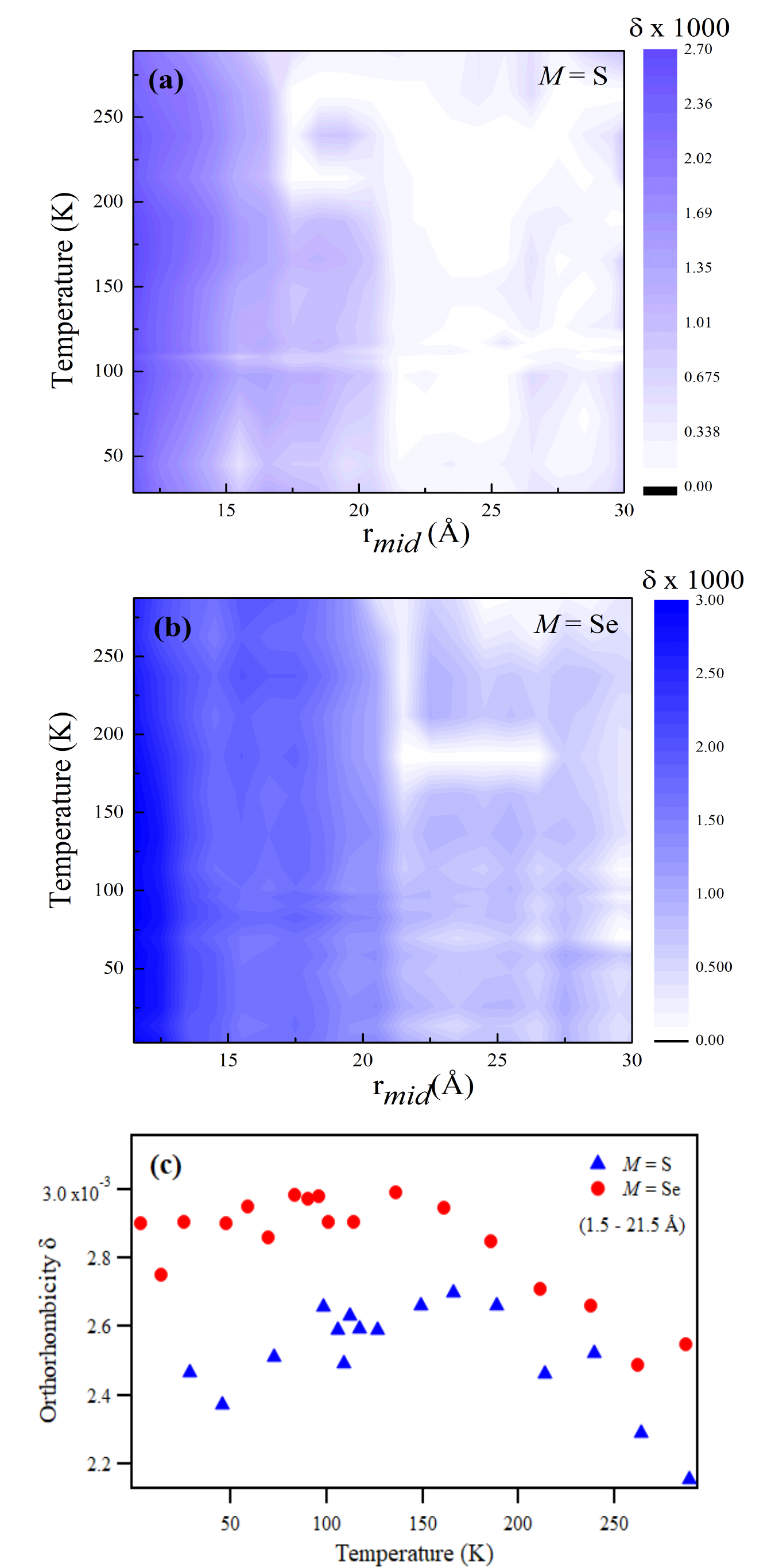}
\caption{Color maps of the refined orthorhombicity determined from neutron PDF analysis for a) $M$ = S and b) $M$ = Se. c) Temperature dependence of short-range (1.5 - 21.5 \AA) orthorhombicity.} 
%At short-ranges over all temperatures, we observe the presence of orthorhombicity for both samples. At large $r$-ranges, the two samples behave differently. In the case of $M$ = S orthorhombicity exists for $r_{mid}$ greater than 35 \r{A} while for $M$ = Se clearly shows tetragonal for distances greater than 35 \r{A}.}  
%Negative values of orthorhombicity indicate that the lattice parameter $b$ is larger than $a$.
\label{fig:color map1}
\end{figure}

%\begin{figure}[!ht]
%\includegraphics[width=3in]{Fig-8.pdf}
%\caption{Color maps of $c$-lattices for {\cSSe} derived from $T$- and $r$-series PDF refinements for a) $M$ = S and b) $M$ = Se.} %Here, the horizontal axis is the midpoint of the fitting ranges $r_{mid}$, the vertical axis is the temperature, and the color bar shows the intensity of $c$-lattice with red (blue) representing high (low) intensity.}
%\label{fig:color map2}
%\end{figure}

\subsection{\label{sec:orthorhombicity}Local Orthorhombicity}
The use of pre-written macros allows \texttt{PDFGUI} to fit data sets through a series of boxcar refinements that differ one from another by corresponding fitting ranges~\cite{farrow2007pdffit2}. Since we have collected PDF data sets over a wide range of temperatures, we used \texttt{PDFGUI} $T$- series and $r$- series  macros to study details of the local to average structure crossover in our materials. These macros allow the setup of boxcar fits, in which the same model is fitted over different real space intervals ($r$- ranges) for different temperatures of the PDF data~\cite{abeykoon2009average}.
 We have performed extensive $T$- and $r$- series PDF fits to determine the atomic structure as a function of temperature and length scale~\cite{PhysRevB.98.180505}. The evolution of the orthorhombicity of the local structure (short-range) to the average structure (long range) is presented in the color maps of Fig.~\ref{fig:color map1}. 
 %The most relevant structural parameter for the local distortion is the orthorhombicity. Color maps of orthorhombicity were prepared, for both $M$ = S and Se by using the results of $T$- and $r$- series refinements based on the orthorhombic symmetry. 
 We did the $T$- series refinement by  performing the PDF fits on low temperature data at first, then we run the sequential PDF fits for all of our data collected at various temperatures.
 For each temperature data, we have performed the $r$- series refinement in a sliding 20 \r{A} data window from [1.5 - 21.5 \r{A}] to [29.5 - 49.5 \r{A}] in 1 \r{A} steps, resulting in 29 fits per temperature. Orthorhombicity ($\delta$) was extracted from all of these fits for all temperatures to produce the color maps. Color maps were made using the \texttt{ORIGIN} software~\cite{origin}. In these color maps, the midpoint of the fitting range r$_{mid}$ is shown on the x-axis, the temperature is shown on the y-axis, and orthorhombicity is shown on the color bar as indicated by the brightness of the colors from light blue to dark blue. The dark blue color denotes high orthorhombicity, light blue represent zero orthorhombicity $i.e.$, tetragonality.  The maps indicate that at short ranges, over all temperatures, the presence of orthorhombicity is observed, but it diminishes over longer length scales. Fig. \ref{fig:color map1}(c) shows the temperature dependence of short-range orthorhombicity for La$_{2}$O$_{2}$Fe$_{2}$O(S, Se)$_{2}$. It was observed that orthorhombicity is independent of temperature below $\sim$ 150 K and at higher temperatures, orthorhombicity decreases slightly with temperature. \\
 %At large $r$-ranges, the two samples behave differently. In the case of $M$ = S, orthorhombicity exists for $r_{mid}$ greater than 35 \r{A} while $M$ = Se clearly shows tetragonality for distances greater than 35 \r{A}.  
 %and dark blue is for negative orthorhombicity indicating the lattice parameter $b$ is greater than $a$. Both $M$ = S and Se, high orthorhombicity is observed in the short-ranges (about 1-3 nm) for all temperatures at which data are collected.
%Moreover, in both $M$ = S and Se plot, we observed negative orthorhombicity, because of the lattice fluctuations lattice parameter $b$ was greater than $a$, at higher ranges within the temperature range from 50K to 205K.\\
\begin{figure}[htp]
\includegraphics[width=2.8in]{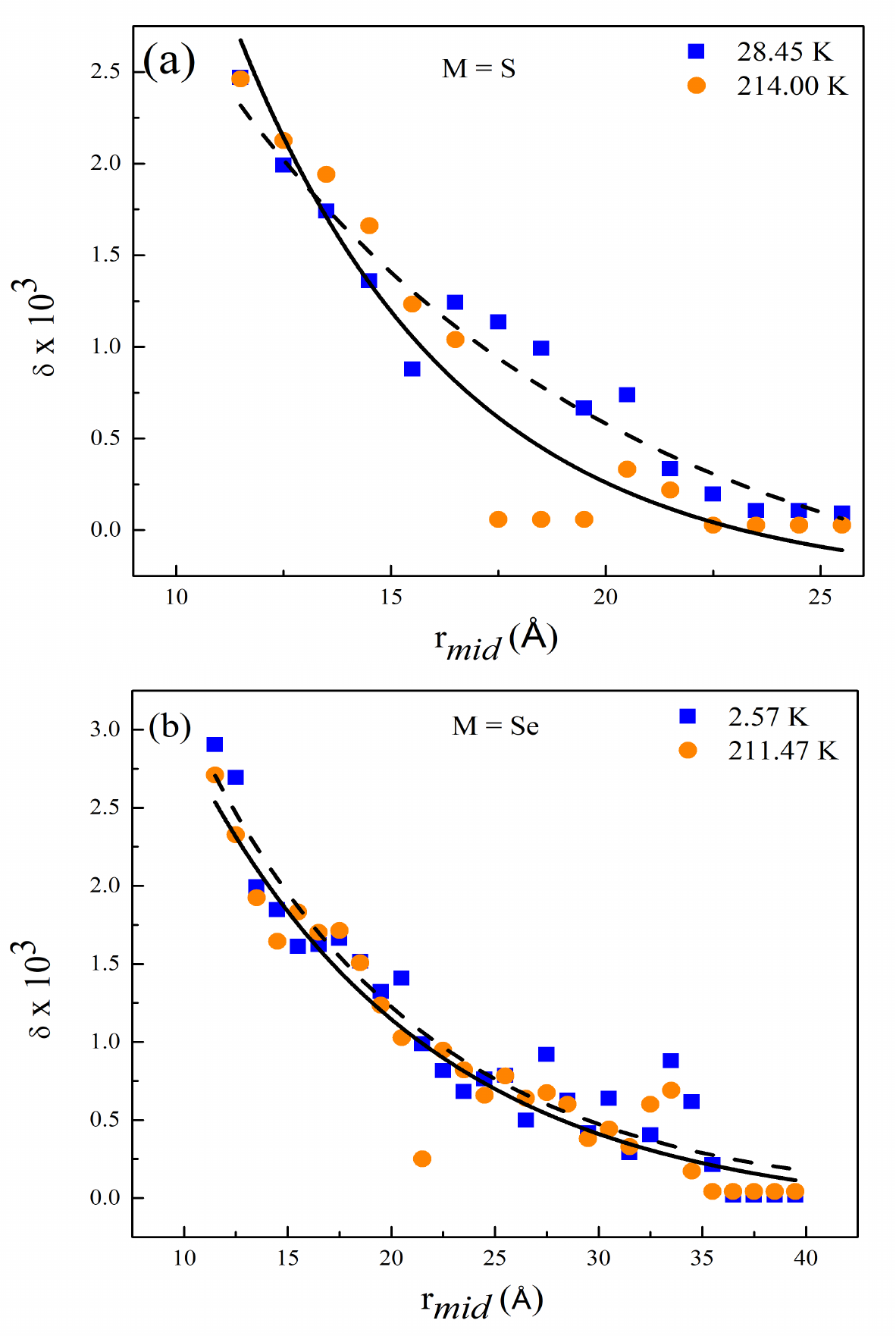}
\caption{Orthorhombicity as a function of fitting range for a) $M$ = S and b) $M$ = Se at low and high temperature.} %Here, the horizontal axis is the midpoint of the fitting range $r_{mid}$, the vertical axis is the temperature, and the color bar shows the intensity of $c$-lattice with red (blue) representing high (low) intensity.}
\label{fig:localortho}
\end{figure}
%\textcolor{red}
Fig. \ref{fig:localortho} presents the local orthorhombicity $\delta$ as a function of the midpoint of the fitting ranges for two representative temperatures.  Data were fitted with exponential decay functions,
\begin{equation}
f(x)=A_{1}{\exp\Big(\frac{-x}{t_{1}}\Big)} + Constant 
\label{equation:E3}
\end{equation}
where decay length is defined as $ln2\times t_{1}$.

Fig. \ref{fig:localortho}(a) shows the orthorhombicity of $M$ = S at 28.45 K and 214.0 K. At both temperatures, the orthorhombicity is largest at short fitting ranges and it asymptotically approaches zero at larger fitting ranges. A black solid (dashed) line shows the exponential fitting function for 28.45 K (214.0 K) with decay lengths of 6.4(2)\r{A} (3.5(9)\r{A}), respectively.  Fig.\ref{fig:localortho}(b) displays the orthorhombicity at different fitting ranges for $M$ = Se at temperature 2.57 K and 211.47 K. Black solid (dashed) line shows the exponential fitting function for 2.57 K (211.47 K) with decay length of 7.5(3)\r{A} (7.9(6)\r{A}), respectively. Figs. \ref{fig:localortho}(a) and (b) provide evidence of the greater orthorhombicity at shorter ranges.
 
%The color maps of the $c$-lattice parameter for $M$ = S, Se determined from $T$- and $r$-series refinement of pair distribution function analysis are presented in Fig. \ref{fig:color map2}. In this figure, the x-axis is the midpoint of fitting range r$_{mid}$, the y-axis is the temperature, and the color bar shows the $c$-lattice parameter with light blue for the high intensity and black for low intensity. $c$-lattice color maps show the evolution of $c$-lattice from the local structure (short-range) to the average structure (long range). For both {\cSSe} materials, we can see an increase in the $c$-lattice as the temperature increases overall $r_{mid}$ ranges.

\section{\label{sec:level4}Discussion}

Neutron powder diffraction experiments confirmed the  average crystal structure of {\cSSe} as tetragonal systems with space group ($I$4/$mmm$)  (see Fig. \ref{Rietveld refinement}) in agreement with the previous studies on these materials \cite{doi:10.1002/anie.199216451, PhysRevB.89.100402,oogarah2018magnetic,PhysRevB.81.214433, PhysRevLett.104.216405, PhysRevB.99.024109}. Unlike many iron pnictides which undergo a structural phase transition from tetragonal to orthorhombic symmetries near $T_s$, {\cSSe} lattice symmetries are not observed to change with temperature. However, both {\cSSe} undergo a magnetic transitions from paramagnetic (PM) phases to antiferromagnetic (AFM) phases at respective N\'{e}el temperatures $T_N$.  %Fig. \ref{fig:phase} compares the phase diagrams in iron pnictides and {\cSSe} Mott insulators.
 Neutron powder diffraction (Fig.~\ref{fig:ADPs} and Table~\ref{tab:RRparameter}) showed that the thermal displacement parameter for the O(2) atom along the $c$-axis was larger than all the other thermal displacement parameters. This was the finding for both materials $M$ = S and Se.  
%Superconducting iron pnictides show nematic fluctuations around $T_s$ that persist to a more ordered phase (nematic phase) until the N\'{e}el temperature in reached.

In order to probe the local structure, we employed the neutron PDF technique to investigate the local changes in the atomic positions and $c$-axes of {\cSSe}.  
%==============================
%\begin{figure}[!h]
%\centering
%\includegraphics[width=0.45\textwidth]{Fig_9_Nov23_all.pdf}
%\caption{(a) A typical iron pnictide superconductor phase diagram with doping versus temperature after references~\cite{Fernandes2014,dagotto2005complexity,dai2012magnetism,li2017nematic}. Schematic phase diagrams of (b) the Mott insulators Ca$_2$O$_2$Fe$_{2.6}$S$_{2-x}$Se$_x$ showing doping versus temperature and (c) the Mott insulators {\cSSe} showing length scale versus temperature.}
%\label{fig:phase}
%\end{figure}
%==================================
%The local crystal structures of {\cSSe} were studied using the pair distribution function technique.  
The PDF analysis revealed similar thermal displacement behavior on short length scales.  In addition, the temperature dependence of the $c$-lattice parameter showed a kink near the N\'{e}el temperature for $M$ = Se (see Fig. \ref{lattice parameter}). However, no discontinuity was observed in the case of $M$ = S. Such a finding is consistent with previous reports of $c$-lattice discontinuities in oxyselenides \cite{PhysRevB.90.165111, PhysRevB.95.174441, PhysRevB.81.214433} and the less prominent occurrence of this behavior in the oxysulfides \cite{oogarah2018magnetic}. The discontinuities were attributed to magnetostrictive effects in the oxyselenides. In the case of oxysulfides, a relatively reduced amount of $c$-lattice discontinuity was argued to be due to shorter $c$-lattices \cite{oogarah2018magnetic}. 
%Discontinuities in the $c$-lattice of oxyselenides were previously reported to be due to magnetostrictive effects \cite{PhysRevB.90.165111, PhysRevB.95.174441, PhysRevB.81.214433}. This behavior is less prominent in oxysulfides due to its shorter $c$-lattice parameter \cite{oogarah2018magnetic}. 
According to Horigane et al.~\cite{horigane2014local} deviations in the $c$-lattice parameter from the average crystallographic structure are due to the fact that the thermal displacement parameter $U_{33}$ for O(2) along the $c$-axis grows rapidly with the increase in temperature. A similar O(2) trend, in SrFeO$_{2}$ \cite{PhysRevLett.112.097001}, was reported to be related to Fe$_2$O planar buckling. %In SrFeO$_{2}$, in-plane O(2) atoms have strong temperature dependence in the $c$-axis direction as seen by high $U_{33}$ values. 

For {\cSSe}, buckling of the Fe$_2$O plane may occur when there is distortion in either the Fe or O(2) sites. In the case of $M$ = Se, a high $U_{33}$ value for O(2) (see Fig. \ref{fig:ADPs}) suggests that the largest distortions should occur for that site. A distortion in the Fe site can lead to two Fe-Se correlations since each Fe atom is surrounded by four Se atoms of the octahedra (see Fig.~\ref{fig:atom_move}b). %Therefore, the modification observed in Fe-Se correlation peak (the third peak of see Fig. \ref{peak splitting}a) might indicate a subtle distortion in the Fe site. 
Overall, these findings for $M$ = Se suggest that there is a distortion either in the Fe or O(2) sites, and both implicated as possible reasons for Fe$_{2}$O buckling.

% Moreover, they discuss the splitting of Fe-Se correlation using the Fe-displacement model. If there is displacement of Fe atom along the $c$-axis then it is possible to have two different Fe-Se correlation. From these discussions they conclude that this discrepancy might be due to the buckling of Fe$_{2}$O plane.

  %Fig.~\ref{fig:phase} compares the phase diagram of iron pnictides (see Fig. \ref{fig:phase}a) and {\cSSe} Mott insulators (see Figs. \ref{fig:phase}b and \ref{fig:phase}c).
  %Study of non-symmetric dynamical fluctuations, which breaks the orientational but not the translational symmetry, is referred as nematic fluctuations \cite{PhysRevB.98.180505,fradkin2010nematic}.The nematic fluctuations, which are deviations from the average structure, may provide the information about the origin of nematicity and its relation to superconductivity mechanism \cite{Kuo958, BOHMER201690, Rosenthal2014, frandsen2019quantitative, koch2019room, PhysRevB.95.024511, RevModPhys.87.855, RevModPhys.83.1589}.
  %It is very crucial to study the local structure of Mott-insulators, in which the long range structural phase transition is not observed, in order to understand superconductivity mechanism. Such studies would further provide the discussion regarding the significance of nematic fluctuations for the superconductivity.
  
  Our analysis of the local structure of \cSSe  reveals the presence of short-range orthorhombic distortions on a length scale of 1 - 2 nm over a temperature range from 2 - 300 K. These distortions are very similar to those found in FeSe~\cite{koch2019room, frandsen2019quantitative} and (Sr, Na)Fe$_{2}$As$_{2}$ ~\cite{PhysRevB.98.180505, PhysRevLett.119.187001}, which have been attributed to short-range, structural nematic fluctuations that persist up to temperatures well above the magnetic and/or structural transition temperatures. Given the structural and magnetic similarities between these iron oxychalcogenides and the superconducting iron pnictides and chalcogenides, we suggest that the local orthorhombicity observed here is likewise related to fluctuating, short-range electronic nematic distortions. The short-range orthorhombicity observed upto high temperature ($\sim$ 300 K) in these systems may reflect a similar large nematic energy scale as was observed in FeSCs such as FeSe \cite{PhysRevX.9.041049} and (Sr, Na)Fe$_{2}$As$_{2}$ \cite{frandsen2019quantitative}.

  %Our analysis of the local structure of {\cSSe} reveals the presence of nematic fluctuations on a length scale of 1 - 2 nm over a temperature range from 2 - 300 K. The observation of nematic fluctuations up to the room temperature is consistent with other pnictide systems. %These findings are consistent with the recent work by Frandsen et al.
 % Our findings are also reminiscent of a recent report of the local structural study of FePn systems which suggests the existence of short-range orthorhombic distortions up to high temperature (500 K) on a length scale of 2 - 3 nm \cite{PhysRevB.98.180505, frandsen2019quantitative, koch2019room, PhysRevLett.119.187001}. 
  
    %Results of our work are very close to those from a prior PDF investigation of Sr$_{1−x}$Na$_{x}$Fe$_2$As$_2$ which suggest the existence of short-range orthorhombic distortions up to high temperature (500 K) on a length scale of 2–3 nm.
  The occurrence of such distortions across a variety of pnictide superconductors and the incoherent electronic systems presented here raise the question of how  are these short-range orthorhombic distortions related to the superconducting mechanism. Inquiry into the role of $electronic$ nematicity has become the key focal point of the field of iron-based high-$T_{c}$ superconductivity.  
  %\textbf{CONSIDER FOR DELETION In most FeSCs, superconductivity is found in close proximity to a magnetically ordered state. However, there is a nematic, non-superconducting state besides magnetism at a specific temperature $T_{nem}$. Nematic ordering has been systematically observed in iron pnictide and chalcogenide superconductors.}
  %At $T_{nem}$, symmetry is broken between x and y directions in the Fe-plane \cite{Fernandes2014}. This reduces the rotational point group symmetry from tetragonal to orthorhombic while preserving the translational symmetry.~\cite{PhysRevB.98.180505,fradkin2010nematic}.
  Studies of short-range orthorhombic fluctuations, such as those reported here, may provide information about the origin of electronic nematicity and its relation to superconductivity mechanism \cite{Kuo958, BOHMER201690, Rosenthal2014, frandsen2019quantitative, koch2019room, PhysRevB.95.024511, RevModPhys.87.855, RevModPhys.83.1589}. %Orthorhombic fluctuations have been shown to drive electronic nematicity in the FeSCs.\cite{PhysRevX.9.041049} 
  
  %The fact that short-range nematic degrees of freedom appear to be active in the Mott insulating, non-superconducting {\cSSe} highlights the ubiquity of structural nematicity in the layered iron-based systems. 
  The presence of local orthorhombicity that appears to be ubiquitous may be suggestive of nematic degrees of freedom being ubiquitously active in layered iron-based systems. It also suggests that active nematic degrees of freedom are not sufficient to guarantee superconductivity, although they may well be necessary. It may be the case that nematic fluctuations and another phase must conspire to produce high-temperature superconductivity in iron-based systems.
  The energy-integrated nature of the PDF measurements means we are probing fluctuations on a time scale of 10$^{-13}$ s or slower. However, we were unable to extract any other temporal information using the PDF data alone \cite{PhysRevLett.119.187001}. Complementary probes such as NMR or M\"ossbauer spectroscopy could help clarify the situation \cite{Freelon2021}.
  %The data presented, both here and in the literature, suggest that the presence of nematic fluctuations does not guarantee the onset of a superconducting phase. It may be the case, in the iron pnictides and iron chalcogenides, that nematic fluctuations, and another phase must conspire to produce high-temperature superconductivity.
  
\section{\label{sec:level5}Conclusions}
The local structure of Mott insulating iron oxychalcogenides was studied using neutron powder diffraction and pair distribution techniques. Neutron powder diffraction showed a similar nuclear structure of $M$ = Se and S with the only difference in the atomic radii of two chalcogens. Pair distribution function analysis indicated the presence of the local distortion between tetragonal and orthorhombic symmetry. These  findings suggest the presence of orthorhombic fluctuations suggestive of short-range nematicity with a typical length scale of 1 - 2 nm in both of these materials.

\section{\label{sec:level6}Acknowledgments}
The authors would like to thank the University of Louisville for funding support. Alfailakawi would like to thank Kuwait University for supporting this work through a graduate scholarship.  Work at University of Houston was supported by the State of Texas through TcSUH. The research in ZJU is supported by the Ministry of Science and Technology of China under Grants No. 2016YFA0300402 and No. 2015CB921004 and the National Natural Science Foundation of China (NSFC) (No. 11974095, 11374261),  and the Fundamental Research Funds for the Central Universities. This work used resources of the Spallation Neutron Source, a DOE office of Science User Facility operated by the Oak Ridge National Laboratory.\\ 
%Notice: This manuscript has been authored by UT-Battelle, LLC, under contract DE-AC05-00OR22725 with the US Department of Energy (DOE). The US government retains and the publisher, by accepting the article for publication, acknowledges that the US government retains a nonexclusive, paid-up, irrevocable, worldwide license to publish or reproduce the published form of this manuscript, or allow others to do so, for US government purposes. DOE will provide public access to these results of federally sponsored research in accordance with the DOE Public Access Plan (http://energy.gov/downloads/doe-public-access-plan).

%\nocite{*}

\bibliography{apssamp}% Produces the bibliography via BibTeX.

\providecommand{\noopsort}[1]{}\providecommand{\singleletter}[1]{#1}%
\begin{thebibliography}{64}%
\makeatletter
\providecommand \@ifxundefined [1]{%
 \@ifx{#1\undefined}
}%
\providecommand \@ifnum [1]{%
 \ifnum #1\expandafter \@firstoftwo
 \else \expandafter \@secondoftwo
 \fi
}%
\providecommand \@ifx [1]{%
 \ifx #1\expandafter \@firstoftwo
 \else \expandafter \@secondoftwo
 \fi
}%
\providecommand \natexlab [1]{#1}%
\providecommand \enquote  [1]{``#1''}%
\providecommand \bibnamefont  [1]{#1}%
\providecommand \bibfnamefont [1]{#1}%
\providecommand \citenamefont [1]{#1}%
\providecommand \href@noop [0]{\@secondoftwo}%
\providecommand \href [0]{\begingroup \@sanitize@url \@href}%
\providecommand \@href[1]{\@@startlink{#1}\@@href}%
\providecommand \@@href[1]{\endgroup#1\@@endlink}%
\providecommand \@sanitize@url [0]{\catcode `\\12\catcode `\$12\catcode
  `\&12\catcode `\#12\catcode `\^12\catcode `\_12\catcode `\%12\relax}%
\providecommand \@@startlink[1]{}%
\providecommand \@@endlink[0]{}%
\providecommand \url  [0]{\begingroup\@sanitize@url \@url }%
\providecommand \@url [1]{\endgroup\@href {#1}{\urlprefix }}%
\providecommand \urlprefix  [0]{URL }%
\providecommand \Eprint [0]{\href }%
\providecommand \doibase [0]{http://dx.doi.org/}%
\providecommand \selectlanguage [0]{\@gobble}%
\providecommand \bibinfo  [0]{\@secondoftwo}%
\providecommand \bibfield  [0]{\@secondoftwo}%
\providecommand \translation [1]{[#1]}%
\providecommand \BibitemOpen [0]{}%
\providecommand \bibitemStop [0]{}%
\providecommand \bibitemNoStop [0]{.\EOS\space}%
\providecommand \EOS [0]{\spacefactor3000\relax}%
\providecommand \BibitemShut  [1]{\csname bibitem#1\endcsname}%
\let\auto@bib@innerbib\@empty
%</preamble>
\bibitem [{\citenamefont {Si}\ \emph {et~al.}(2016)\citenamefont {Si},
  \citenamefont {Yu},\ and\ \citenamefont {Abrahams}}]{Si2016}%
  \BibitemOpen
  \bibfield  {author} {\bibinfo {author} {\bibfnamefont {Q.}~\bibnamefont
  {Si}}, \bibinfo {author} {\bibfnamefont {R.}~\bibnamefont {Yu}}, \ and\
  \bibinfo {author} {\bibfnamefont {E.}~\bibnamefont {Abrahams}},\ }\href
  {https://doi.org/10.1038/natrevmats.2016.17} {\bibfield  {journal} {\bibinfo
  {journal} {Nature Reviews Materials}\ }\textbf {\bibinfo {volume} {1}},\
  \bibinfo {pages} {16017} (\bibinfo {year} {2016})}\BibitemShut {NoStop}%
\bibitem [{\citenamefont {Paglione}\ and\ \citenamefont
  {Greene}(2010)}]{Paglione2010}%
  \BibitemOpen
  \bibfield  {author} {\bibinfo {author} {\bibfnamefont {J.}~\bibnamefont
  {Paglione}}\ and\ \bibinfo {author} {\bibfnamefont {R.~L.}\ \bibnamefont
  {Greene}},\ }\href {https://doi.org/10.1038/nphys1759} {\bibfield  {journal}
  {\bibinfo  {journal} {Nature Physics}\ }\textbf {\bibinfo {volume} {6}},\
  \bibinfo {pages} {645} (\bibinfo {year} {2010})}\BibitemShut {NoStop}%
\bibitem [{\citenamefont {Zhao}\ \emph {et~al.}(2019)\citenamefont {Zhao},
  \citenamefont {Wang}, \citenamefont {Huang}, \citenamefont {Wu},
  \citenamefont {Sun}, \citenamefont {Fan}, \citenamefont {Song}, \citenamefont
  {Jin},\ and\ \citenamefont {Chen}}]{zhao2019structural}%
  \BibitemOpen
  \bibfield  {author} {\bibinfo {author} {\bibfnamefont {L.}~\bibnamefont
  {Zhao}}, \bibinfo {author} {\bibfnamefont {D.}~\bibnamefont {Wang}}, \bibinfo
  {author} {\bibfnamefont {Q.}~\bibnamefont {Huang}}, \bibinfo {author}
  {\bibfnamefont {H.}~\bibnamefont {Wu}}, \bibinfo {author} {\bibfnamefont
  {R.}~\bibnamefont {Sun}}, \bibinfo {author} {\bibfnamefont {X.}~\bibnamefont
  {Fan}}, \bibinfo {author} {\bibfnamefont {Y.}~\bibnamefont {Song}}, \bibinfo
  {author} {\bibfnamefont {S.}~\bibnamefont {Jin}}, \ and\ \bibinfo {author}
  {\bibfnamefont {X.}~\bibnamefont {Chen}},\ }\href@noop {} {\bibfield
  {journal} {\bibinfo  {journal} {Physical Review B}\ }\textbf {\bibinfo
  {volume} {99}},\ \bibinfo {pages} {094503} (\bibinfo {year}
  {2019})}\BibitemShut {NoStop}%
\bibitem [{\citenamefont {Yildirim}(2009)}]{PhysRevLett.102.037003}%
  \BibitemOpen
  \bibfield  {author} {\bibinfo {author} {\bibfnamefont {T.}~\bibnamefont
  {Yildirim}},\ }\href {\doibase 10.1103/PhysRevLett.102.037003} {\bibfield
  {journal} {\bibinfo  {journal} {Phys. Rev. Lett.}\ }\textbf {\bibinfo
  {volume} {102}},\ \bibinfo {pages} {037003} (\bibinfo {year}
  {2009})}\BibitemShut {NoStop}%
\bibitem [{\citenamefont {Si}\ and\ \citenamefont
  {Abrahams}(2008)}]{PhysRevLett.101.076401}%
  \BibitemOpen
  \bibfield  {author} {\bibinfo {author} {\bibfnamefont {Q.}~\bibnamefont
  {Si}}\ and\ \bibinfo {author} {\bibfnamefont {E.}~\bibnamefont {Abrahams}},\
  }\href {\doibase 10.1103/PhysRevLett.101.076401} {\bibfield  {journal}
  {\bibinfo  {journal} {Phys. Rev. Lett.}\ }\textbf {\bibinfo {volume} {101}},\
  \bibinfo {pages} {076401} (\bibinfo {year} {2008})}\BibitemShut {NoStop}%
\bibitem [{\citenamefont {Mansart}\ \emph {et~al.}(2010)\citenamefont
  {Mansart}, \citenamefont {Boschetto}, \citenamefont {Savoia}, \citenamefont
  {Rullier-Albenque}, \citenamefont {Bouquet}, \citenamefont {Papalazarou},
  \citenamefont {Forget}, \citenamefont {Colson}, \citenamefont {Rousse},\ and\
  \citenamefont {Marsi}}]{PhysRevB.82.024513}%
  \BibitemOpen
  \bibfield  {author} {\bibinfo {author} {\bibfnamefont {B.}~\bibnamefont
  {Mansart}}, \bibinfo {author} {\bibfnamefont {D.}~\bibnamefont {Boschetto}},
  \bibinfo {author} {\bibfnamefont {A.}~\bibnamefont {Savoia}}, \bibinfo
  {author} {\bibfnamefont {F.}~\bibnamefont {Rullier-Albenque}}, \bibinfo
  {author} {\bibfnamefont {F.}~\bibnamefont {Bouquet}}, \bibinfo {author}
  {\bibfnamefont {E.}~\bibnamefont {Papalazarou}}, \bibinfo {author}
  {\bibfnamefont {A.}~\bibnamefont {Forget}}, \bibinfo {author} {\bibfnamefont
  {D.}~\bibnamefont {Colson}}, \bibinfo {author} {\bibfnamefont
  {A.}~\bibnamefont {Rousse}}, \ and\ \bibinfo {author} {\bibfnamefont
  {M.}~\bibnamefont {Marsi}},\ }\href {\doibase 10.1103/PhysRevB.82.024513}
  {\bibfield  {journal} {\bibinfo  {journal} {Phys. Rev. B}\ }\textbf {\bibinfo
  {volume} {82}},\ \bibinfo {pages} {024513} (\bibinfo {year}
  {2010})}\BibitemShut {NoStop}%
\bibitem [{\citenamefont {Lee}\ \emph {et~al.}(2019)\citenamefont {Lee},
  \citenamefont {Roh}, \citenamefont {Seo}, \citenamefont {Lee}, \citenamefont
  {Jung}, \citenamefont {Ok}, \citenamefont {Jung}, \citenamefont {Kang},
  \citenamefont {Lee}, \citenamefont {Kim} \emph
  {et~al.}}]{lee2019temperature}%
  \BibitemOpen
  \bibfield  {author} {\bibinfo {author} {\bibfnamefont {S.}~\bibnamefont
  {Lee}}, \bibinfo {author} {\bibfnamefont {S.}~\bibnamefont {Roh}}, \bibinfo
  {author} {\bibfnamefont {Y.-S.}\ \bibnamefont {Seo}}, \bibinfo {author}
  {\bibfnamefont {M.}~\bibnamefont {Lee}}, \bibinfo {author} {\bibfnamefont
  {E.}~\bibnamefont {Jung}}, \bibinfo {author} {\bibfnamefont {J.~M.}\
  \bibnamefont {Ok}}, \bibinfo {author} {\bibfnamefont {M.-C.}\ \bibnamefont
  {Jung}}, \bibinfo {author} {\bibfnamefont {B.}~\bibnamefont {Kang}}, \bibinfo
  {author} {\bibfnamefont {K.-W.}\ \bibnamefont {Lee}}, \bibinfo {author}
  {\bibfnamefont {J.~S.}\ \bibnamefont {Kim}},  \emph {et~al.},\ }\href@noop {}
  {\bibfield  {journal} {\bibinfo  {journal} {Journal of Physics: Condensed
  Matter}\ }\textbf {\bibinfo {volume} {31}},\ \bibinfo {pages} {445602}
  (\bibinfo {year} {2019})}\BibitemShut {NoStop}%
\bibitem [{\citenamefont {Wong}\ and\ \citenamefont
  {Lortz}(2019)}]{wong2019antiferromagnetic}%
  \BibitemOpen
  \bibfield  {author} {\bibinfo {author} {\bibfnamefont {C.~H.}\ \bibnamefont
  {Wong}}\ and\ \bibinfo {author} {\bibfnamefont {R.}~\bibnamefont {Lortz}},\
  }\href@noop {} {\bibfield  {journal} {\bibinfo  {journal} {arXiv preprint
  arXiv:1902.06463}\ } (\bibinfo {year} {2019})}\BibitemShut {NoStop}%
\bibitem [{\citenamefont {Abrahams}\ and\ \citenamefont
  {Si}(2011)}]{abrahams2011quantum}%
  \BibitemOpen
  \bibfield  {author} {\bibinfo {author} {\bibfnamefont {E.}~\bibnamefont
  {Abrahams}}\ and\ \bibinfo {author} {\bibfnamefont {Q.}~\bibnamefont {Si}},\
  }\href@noop {} {\bibfield  {journal} {\bibinfo  {journal} {Journal of
  physics: Condensed matter}\ }\textbf {\bibinfo {volume} {23}},\ \bibinfo
  {pages} {223201} (\bibinfo {year} {2011})}\BibitemShut {NoStop}%
\bibitem [{\citenamefont {Shrivastava}\ and\ \citenamefont
  {Kumar}(2019)}]{shrivastava2019high}%
  \BibitemOpen
  \bibfield  {author} {\bibinfo {author} {\bibfnamefont {S.~K.}\ \bibnamefont
  {Shrivastava}}\ and\ \bibinfo {author} {\bibfnamefont {G.}~\bibnamefont
  {Kumar}},\ }\href@noop {} {\bibfield  {journal} {\bibinfo  {journal}
  {International Journal of Emerging Technologies and Innovative Research}\
  }\textbf {\bibinfo {volume} {6}},\ \bibinfo {pages} {417} (\bibinfo {year}
  {2019})}\BibitemShut {NoStop}%
\bibitem [{\citenamefont {Landsgesell}\ \emph {et~al.}(2014)\citenamefont
  {Landsgesell}, \citenamefont {Proke{\v{s}}}, \citenamefont {Hansen},\ and\
  \citenamefont {Frontzek}}]{landsgesell2014unexpected}%
  \BibitemOpen
  \bibfield  {author} {\bibinfo {author} {\bibfnamefont {S.}~\bibnamefont
  {Landsgesell}}, \bibinfo {author} {\bibfnamefont {K.}~\bibnamefont
  {Proke{\v{s}}}}, \bibinfo {author} {\bibfnamefont {T.}~\bibnamefont
  {Hansen}}, \ and\ \bibinfo {author} {\bibfnamefont {M.}~\bibnamefont
  {Frontzek}},\ }\href@noop {} {\bibfield  {journal} {\bibinfo  {journal} {Acta
  Materialia}\ }\textbf {\bibinfo {volume} {66}},\ \bibinfo {pages} {232}
  (\bibinfo {year} {2014})}\BibitemShut {NoStop}%
\bibitem [{\citenamefont {Zhu}\ \emph {et~al.}(2010)\citenamefont {Zhu},
  \citenamefont {Yu}, \citenamefont {Wang}, \citenamefont {Zhao}, \citenamefont
  {Jones}, \citenamefont {Dai}, \citenamefont {Abrahams}, \citenamefont
  {Morosan}, \citenamefont {Fang},\ and\ \citenamefont
  {Si}}]{PhysRevLett.104.216405}%
  \BibitemOpen
  \bibfield  {author} {\bibinfo {author} {\bibfnamefont {J.-X.}\ \bibnamefont
  {Zhu}}, \bibinfo {author} {\bibfnamefont {R.}~\bibnamefont {Yu}}, \bibinfo
  {author} {\bibfnamefont {H.}~\bibnamefont {Wang}}, \bibinfo {author}
  {\bibfnamefont {L.~L.}\ \bibnamefont {Zhao}}, \bibinfo {author}
  {\bibfnamefont {M.~D.}\ \bibnamefont {Jones}}, \bibinfo {author}
  {\bibfnamefont {J.}~\bibnamefont {Dai}}, \bibinfo {author} {\bibfnamefont
  {E.}~\bibnamefont {Abrahams}}, \bibinfo {author} {\bibfnamefont
  {E.}~\bibnamefont {Morosan}}, \bibinfo {author} {\bibfnamefont
  {M.}~\bibnamefont {Fang}}, \ and\ \bibinfo {author} {\bibfnamefont
  {Q.}~\bibnamefont {Si}},\ }\href {\doibase 10.1103/PhysRevLett.104.216405}
  {\bibfield  {journal} {\bibinfo  {journal} {Phys. Rev. Lett.}\ }\textbf
  {\bibinfo {volume} {104}},\ \bibinfo {pages} {216405} (\bibinfo {year}
  {2010})}\BibitemShut {NoStop}%
\bibitem [{\citenamefont {Lee}\ \emph {et~al.}(2006)\citenamefont {Lee},
  \citenamefont {Nagaosa},\ and\ \citenamefont {Wen}}]{RevModPhys.78.17}%
  \BibitemOpen
  \bibfield  {author} {\bibinfo {author} {\bibfnamefont {P.~A.}\ \bibnamefont
  {Lee}}, \bibinfo {author} {\bibfnamefont {N.}~\bibnamefont {Nagaosa}}, \ and\
  \bibinfo {author} {\bibfnamefont {X.-G.}\ \bibnamefont {Wen}},\ }\href
  {\doibase 10.1103/RevModPhys.78.17} {\bibfield  {journal} {\bibinfo
  {journal} {Rev. Mod. Phys.}\ }\textbf {\bibinfo {volume} {78}},\ \bibinfo
  {pages} {17} (\bibinfo {year} {2006})}\BibitemShut {NoStop}%
\bibitem [{\citenamefont {Fischer}\ \emph {et~al.}(2014)\citenamefont
  {Fischer}, \citenamefont {Wu}, \citenamefont {Lawler}, \citenamefont
  {Paramekanti},\ and\ \citenamefont {Kim}}]{Fischer_2014}%
  \BibitemOpen
  \bibfield  {author} {\bibinfo {author} {\bibfnamefont {M.~H.}\ \bibnamefont
  {Fischer}}, \bibinfo {author} {\bibfnamefont {S.}~\bibnamefont {Wu}},
  \bibinfo {author} {\bibfnamefont {M.}~\bibnamefont {Lawler}}, \bibinfo
  {author} {\bibfnamefont {A.}~\bibnamefont {Paramekanti}}, \ and\ \bibinfo
  {author} {\bibfnamefont {E.-A.}\ \bibnamefont {Kim}},\ }\href {\doibase
  10.1088/1367-2630/16/9/093057} {\bibfield  {journal} {\bibinfo  {journal}
  {New Journal of Physics}\ }\textbf {\bibinfo {volume} {16}},\ \bibinfo
  {pages} {093057} (\bibinfo {year} {2014})}\BibitemShut {NoStop}%
\bibitem [{\citenamefont {Mayer}\ \emph {et~al.}(1992)\citenamefont {Mayer},
  \citenamefont {Schneemeyer}, \citenamefont {Siegrist}, \citenamefont
  {Waszczak},\ and\ \citenamefont {Van~Dover}}]{doi:10.1002/anie.199216451}%
  \BibitemOpen
  \bibfield  {author} {\bibinfo {author} {\bibfnamefont {J.~M.}\ \bibnamefont
  {Mayer}}, \bibinfo {author} {\bibfnamefont {L.~F.}\ \bibnamefont
  {Schneemeyer}}, \bibinfo {author} {\bibfnamefont {T.}~\bibnamefont
  {Siegrist}}, \bibinfo {author} {\bibfnamefont {J.~V.}\ \bibnamefont
  {Waszczak}}, \ and\ \bibinfo {author} {\bibfnamefont {B.}~\bibnamefont
  {Van~Dover}},\ }\href {\doibase 10.1002/anie.199216451} {\bibfield  {journal}
  {\bibinfo  {journal} {Angewandte Chemie International Edition in English}\
  }\textbf {\bibinfo {volume} {31}},\ \bibinfo {pages} {1645} (\bibinfo {year}
  {1992})}\BibitemShut {NoStop}%
\bibitem [{\citenamefont {Freelon}\ \emph {et~al.}(2019)\citenamefont
  {Freelon}, \citenamefont {Yamani}, \citenamefont {Swainson}, \citenamefont
  {Flacau}, \citenamefont {Karki}, \citenamefont {Liu}, \citenamefont {Craco},
  \citenamefont {Laad}, \citenamefont {Wang}, \citenamefont {Chen},
  \citenamefont {Birgeneau},\ and\ \citenamefont {Fang}}]{PhysRevB.99.024109}%
  \BibitemOpen
  \bibfield  {author} {\bibinfo {author} {\bibfnamefont {B.}~\bibnamefont
  {Freelon}}, \bibinfo {author} {\bibfnamefont {Z.}~\bibnamefont {Yamani}},
  \bibinfo {author} {\bibfnamefont {I.}~\bibnamefont {Swainson}}, \bibinfo
  {author} {\bibfnamefont {R.}~\bibnamefont {Flacau}}, \bibinfo {author}
  {\bibfnamefont {B.}~\bibnamefont {Karki}}, \bibinfo {author} {\bibfnamefont
  {Y.~H.}\ \bibnamefont {Liu}}, \bibinfo {author} {\bibfnamefont
  {L.}~\bibnamefont {Craco}}, \bibinfo {author} {\bibfnamefont {M.~S.}\
  \bibnamefont {Laad}}, \bibinfo {author} {\bibfnamefont {M.}~\bibnamefont
  {Wang}}, \bibinfo {author} {\bibfnamefont {J.}~\bibnamefont {Chen}}, \bibinfo
  {author} {\bibfnamefont {R.~J.}\ \bibnamefont {Birgeneau}}, \ and\ \bibinfo
  {author} {\bibfnamefont {M.}~\bibnamefont {Fang}},\ }\href {\doibase
  10.1103/PhysRevB.99.024109} {\bibfield  {journal} {\bibinfo  {journal} {Phys.
  Rev. B}\ }\textbf {\bibinfo {volume} {99}},\ \bibinfo {pages} {024109}
  (\bibinfo {year} {2019})}\BibitemShut {NoStop}%
\bibitem [{\citenamefont {McCabe}\ \emph
  {et~al.}(2014{\natexlab{a}})\citenamefont {McCabe}, \citenamefont {Stock},
  \citenamefont {Rodriguez}, \citenamefont {Wills}, \citenamefont {Taylor},\
  and\ \citenamefont {Evans}}]{PhysRevB.89.100402}%
  \BibitemOpen
  \bibfield  {author} {\bibinfo {author} {\bibfnamefont {E.~E.}\ \bibnamefont
  {McCabe}}, \bibinfo {author} {\bibfnamefont {C.}~\bibnamefont {Stock}},
  \bibinfo {author} {\bibfnamefont {E.~E.}\ \bibnamefont {Rodriguez}}, \bibinfo
  {author} {\bibfnamefont {A.~S.}\ \bibnamefont {Wills}}, \bibinfo {author}
  {\bibfnamefont {J.~W.}\ \bibnamefont {Taylor}}, \ and\ \bibinfo {author}
  {\bibfnamefont {J.~S.~O.}\ \bibnamefont {Evans}},\ }\href {\doibase
  10.1103/PhysRevB.89.100402} {\bibfield  {journal} {\bibinfo  {journal} {Phys.
  Rev. B}\ }\textbf {\bibinfo {volume} {89}},\ \bibinfo {pages} {100402(R)}
  (\bibinfo {year} {2014}{\natexlab{a}})}\BibitemShut {NoStop}%
\bibitem [{\citenamefont {Oogarah}\ \emph {et~al.}(2018)\citenamefont
  {Oogarah}, \citenamefont {Suard},\ and\ \citenamefont
  {McCabe}}]{oogarah2018magnetic}%
  \BibitemOpen
  \bibfield  {author} {\bibinfo {author} {\bibfnamefont {R.~K.}\ \bibnamefont
  {Oogarah}}, \bibinfo {author} {\bibfnamefont {E.}~\bibnamefont {Suard}}, \
  and\ \bibinfo {author} {\bibfnamefont {E.~E.}\ \bibnamefont {McCabe}},\
  }\href@noop {} {\bibfield  {journal} {\bibinfo  {journal} {Journal of
  Magnetism and Magnetic Materials}\ }\textbf {\bibinfo {volume} {446}},\
  \bibinfo {pages} {101} (\bibinfo {year} {2018})}\BibitemShut {NoStop}%
\bibitem [{\citenamefont {Free}\ and\ \citenamefont
  {Evans}(2010)}]{PhysRevB.81.214433}%
  \BibitemOpen
  \bibfield  {author} {\bibinfo {author} {\bibfnamefont {D.~G.}\ \bibnamefont
  {Free}}\ and\ \bibinfo {author} {\bibfnamefont {J.~S.~O.}\ \bibnamefont
  {Evans}},\ }\href {\doibase 10.1103/PhysRevB.81.214433} {\bibfield  {journal}
  {\bibinfo  {journal} {Phys. Rev. B}\ }\textbf {\bibinfo {volume} {81}},\
  \bibinfo {pages} {214433} (\bibinfo {year} {2010})}\BibitemShut {NoStop}%
\bibitem [{\citenamefont {Craco}\ \emph {et~al.}(2018)\citenamefont {Craco},
  \citenamefont {Freelon}, \citenamefont {Alafailakawi}, \citenamefont
  {Karki},\ and\ \citenamefont {Leoni}}]{PhysRevB.98.045130}%
  \BibitemOpen
  \bibfield  {author} {\bibinfo {author} {\bibfnamefont {L.}~\bibnamefont
  {Craco}}, \bibinfo {author} {\bibfnamefont {B.}~\bibnamefont {Freelon}},
  \bibinfo {author} {\bibfnamefont {A.~M.}\ \bibnamefont {Alafailakawi}},
  \bibinfo {author} {\bibfnamefont {B.}~\bibnamefont {Karki}}, \ and\ \bibinfo
  {author} {\bibfnamefont {S.}~\bibnamefont {Leoni}},\ }\href {\doibase
  10.1103/PhysRevB.98.045130} {\bibfield  {journal} {\bibinfo  {journal} {Phys.
  Rev. B}\ }\textbf {\bibinfo {volume} {98}},\ \bibinfo {pages} {045130}
  (\bibinfo {year} {2018})}\BibitemShut {NoStop}%
\bibitem [{\citenamefont {Stock}\ and\ \citenamefont
  {McCabe}(2016)}]{Stock_2016}%
  \BibitemOpen
  \bibfield  {author} {\bibinfo {author} {\bibfnamefont {C.}~\bibnamefont
  {Stock}}\ and\ \bibinfo {author} {\bibfnamefont {E.~E.}\ \bibnamefont
  {McCabe}},\ }\href {\doibase 10.1088/0953-8984/28/45/453001} {\bibfield
  {journal} {\bibinfo  {journal} {Journal of Physics: Condensed Matter}\
  }\textbf {\bibinfo {volume} {28}},\ \bibinfo {pages} {453001} (\bibinfo
  {year} {2016})}\BibitemShut {NoStop}%
\bibitem [{\citenamefont {G\"unther}\ \emph {et~al.}(2014)\citenamefont
  {G\"unther}, \citenamefont {Kamusella}, \citenamefont {Sarkar}, \citenamefont
  {Goltz}, \citenamefont {Luetkens}, \citenamefont {Pascua}, \citenamefont
  {Do}, \citenamefont {Choi}, \citenamefont {Zhou}, \citenamefont {Blum},
  \citenamefont {Wurmehl}, \citenamefont {B\"uchner},\ and\ \citenamefont
  {Klauss}}]{PhysRevB.90.184408}%
  \BibitemOpen
  \bibfield  {author} {\bibinfo {author} {\bibfnamefont {M.}~\bibnamefont
  {G\"unther}}, \bibinfo {author} {\bibfnamefont {S.}~\bibnamefont
  {Kamusella}}, \bibinfo {author} {\bibfnamefont {R.}~\bibnamefont {Sarkar}},
  \bibinfo {author} {\bibfnamefont {T.}~\bibnamefont {Goltz}}, \bibinfo
  {author} {\bibfnamefont {H.}~\bibnamefont {Luetkens}}, \bibinfo {author}
  {\bibfnamefont {G.}~\bibnamefont {Pascua}}, \bibinfo {author} {\bibfnamefont
  {S.-H.}\ \bibnamefont {Do}}, \bibinfo {author} {\bibfnamefont {K.-Y.}\
  \bibnamefont {Choi}}, \bibinfo {author} {\bibfnamefont {H.~D.}\ \bibnamefont
  {Zhou}}, \bibinfo {author} {\bibfnamefont {C.~G.~F.}\ \bibnamefont {Blum}},
  \bibinfo {author} {\bibfnamefont {S.}~\bibnamefont {Wurmehl}}, \bibinfo
  {author} {\bibfnamefont {B.}~\bibnamefont {B\"uchner}}, \ and\ \bibinfo
  {author} {\bibfnamefont {H.-H.}\ \bibnamefont {Klauss}},\ }\href {\doibase
  10.1103/PhysRevB.90.184408} {\bibfield  {journal} {\bibinfo  {journal} {Phys.
  Rev. B}\ }\textbf {\bibinfo {volume} {90}},\ \bibinfo {pages} {184408}
  (\bibinfo {year} {2014})}\BibitemShut {NoStop}%
\bibitem [{\citenamefont {Dai}\ \emph {et~al.}(2012)\citenamefont {Dai},
  \citenamefont {Hu},\ and\ \citenamefont {Dagotto}}]{dai2012magnetism}%
  \BibitemOpen
  \bibfield  {author} {\bibinfo {author} {\bibfnamefont {P.}~\bibnamefont
  {Dai}}, \bibinfo {author} {\bibfnamefont {J.}~\bibnamefont {Hu}}, \ and\
  \bibinfo {author} {\bibfnamefont {E.}~\bibnamefont {Dagotto}},\ }\href@noop
  {} {\bibfield  {journal} {\bibinfo  {journal} {Nature Physics}\ }\textbf
  {\bibinfo {volume} {8}},\ \bibinfo {pages} {709} (\bibinfo {year}
  {2012})}\BibitemShut {NoStop}%
\bibitem [{\citenamefont {Fernandes}\ \emph {et~al.}(2014)\citenamefont
  {Fernandes}, \citenamefont {Chubukov},\ and\ \citenamefont
  {Schmalian}}]{Fernandes2014}%
  \BibitemOpen
  \bibfield  {author} {\bibinfo {author} {\bibfnamefont {R.~M.}\ \bibnamefont
  {Fernandes}}, \bibinfo {author} {\bibfnamefont {A.~V.}\ \bibnamefont
  {Chubukov}}, \ and\ \bibinfo {author} {\bibfnamefont {J.}~\bibnamefont
  {Schmalian}},\ }\href {https://doi.org/10.1038/nphys2877} {\bibfield
  {journal} {\bibinfo  {journal} {Nature Physics}\ }\textbf {\bibinfo {volume}
  {10}},\ \bibinfo {pages} {97} (\bibinfo {year} {2014})}\BibitemShut {NoStop}%
\bibitem [{\citenamefont {Chu}\ \emph {et~al.}(2012)\citenamefont {Chu},
  \citenamefont {Kuo}, \citenamefont {Analytis},\ and\ \citenamefont
  {Fisher}}]{chu2012divergent}%
  \BibitemOpen
  \bibfield  {author} {\bibinfo {author} {\bibfnamefont {J.-H.}\ \bibnamefont
  {Chu}}, \bibinfo {author} {\bibfnamefont {H.-H.}\ \bibnamefont {Kuo}},
  \bibinfo {author} {\bibfnamefont {J.~G.}\ \bibnamefont {Analytis}}, \ and\
  \bibinfo {author} {\bibfnamefont {I.~R.}\ \bibnamefont {Fisher}},\
  }\href@noop {} {\bibfield  {journal} {\bibinfo  {journal} {Science}\ }\textbf
  {\bibinfo {volume} {337}},\ \bibinfo {pages} {710} (\bibinfo {year}
  {2012})}\BibitemShut {NoStop}%
\bibitem [{\citenamefont {Li}\ \emph {et~al.}(2017)\citenamefont {Li},
  \citenamefont {Pereira}, \citenamefont {Yuan}, \citenamefont {Lv},
  \citenamefont {Jiang}, \citenamefont {Lu}, \citenamefont {Lin}, \citenamefont
  {Liu}, \citenamefont {Wang}, \citenamefont {Li} \emph
  {et~al.}}]{li2017nematic}%
  \BibitemOpen
  \bibfield  {author} {\bibinfo {author} {\bibfnamefont {J.}~\bibnamefont
  {Li}}, \bibinfo {author} {\bibfnamefont {P.~J.}\ \bibnamefont {Pereira}},
  \bibinfo {author} {\bibfnamefont {J.}~\bibnamefont {Yuan}}, \bibinfo {author}
  {\bibfnamefont {Y.-Y.}\ \bibnamefont {Lv}}, \bibinfo {author} {\bibfnamefont
  {M.-P.}\ \bibnamefont {Jiang}}, \bibinfo {author} {\bibfnamefont
  {D.}~\bibnamefont {Lu}}, \bibinfo {author} {\bibfnamefont {Z.-Q.}\
  \bibnamefont {Lin}}, \bibinfo {author} {\bibfnamefont {Y.-J.}\ \bibnamefont
  {Liu}}, \bibinfo {author} {\bibfnamefont {J.-F.}\ \bibnamefont {Wang}},
  \bibinfo {author} {\bibfnamefont {L.}~\bibnamefont {Li}},  \emph {et~al.},\
  }\href@noop {} {\bibfield  {journal} {\bibinfo  {journal} {Nature
  communications}\ }\textbf {\bibinfo {volume} {8}},\ \bibinfo {pages} {1880}
  (\bibinfo {year} {2017})}\BibitemShut {NoStop}%
\bibitem [{\citenamefont {Fernandes}\ and\ \citenamefont
  {Millis}(2013)}]{fernandes2013nematicity}%
  \BibitemOpen
  \bibfield  {author} {\bibinfo {author} {\bibfnamefont {R.~M.}\ \bibnamefont
  {Fernandes}}\ and\ \bibinfo {author} {\bibfnamefont {A.~J.}\ \bibnamefont
  {Millis}},\ }\href@noop {} {\bibfield  {journal} {\bibinfo  {journal}
  {Physical review letters}\ }\textbf {\bibinfo {volume} {111}},\ \bibinfo
  {pages} {127001} (\bibinfo {year} {2013})}\BibitemShut {NoStop}%
\bibitem [{\citenamefont {Fradkin}\ \emph {et~al.}(2010)\citenamefont
  {Fradkin}, \citenamefont {Kivelson}, \citenamefont {Lawler}, \citenamefont
  {Eisenstein},\ and\ \citenamefont {Mackenzie}}]{fradkin2010nematic}%
  \BibitemOpen
  \bibfield  {author} {\bibinfo {author} {\bibfnamefont {E.}~\bibnamefont
  {Fradkin}}, \bibinfo {author} {\bibfnamefont {S.~A.}\ \bibnamefont
  {Kivelson}}, \bibinfo {author} {\bibfnamefont {M.~J.}\ \bibnamefont
  {Lawler}}, \bibinfo {author} {\bibfnamefont {J.~P.}\ \bibnamefont
  {Eisenstein}}, \ and\ \bibinfo {author} {\bibfnamefont {A.~P.}\ \bibnamefont
  {Mackenzie}},\ }\href@noop {} {\bibfield  {journal} {\bibinfo  {journal}
  {Annu. Rev. Condens. Matter Phys.}\ }\textbf {\bibinfo {volume} {1}},\
  \bibinfo {pages} {153} (\bibinfo {year} {2010})}\BibitemShut {NoStop}%
\bibitem [{\citenamefont {Chubukov}\ \emph {et~al.}(2015)\citenamefont
  {Chubukov}, \citenamefont {Fernandes},\ and\ \citenamefont
  {Schmalian}}]{chubukov2015origin}%
  \BibitemOpen
  \bibfield  {author} {\bibinfo {author} {\bibfnamefont {A.~V.}\ \bibnamefont
  {Chubukov}}, \bibinfo {author} {\bibfnamefont {R.~M.}\ \bibnamefont
  {Fernandes}}, \ and\ \bibinfo {author} {\bibfnamefont {J.}~\bibnamefont
  {Schmalian}},\ }\href@noop {} {\bibfield  {journal} {\bibinfo  {journal}
  {Physical Review B}\ }\textbf {\bibinfo {volume} {91}},\ \bibinfo {pages}
  {201105(R)} (\bibinfo {year} {2015})}\BibitemShut {NoStop}%
\bibitem [{\citenamefont {Frandsen}\ \emph {et~al.}(2018)\citenamefont
  {Frandsen}, \citenamefont {Taddei}, \citenamefont {Bugaris}, \citenamefont
  {Stadel}, \citenamefont {Yi}, \citenamefont {Acharya}, \citenamefont
  {Osborn}, \citenamefont {Rosenkranz}, \citenamefont {Chmaissem},\ and\
  \citenamefont {Birgeneau}}]{PhysRevB.98.180505}%
  \BibitemOpen
  \bibfield  {author} {\bibinfo {author} {\bibfnamefont {B.~A.}\ \bibnamefont
  {Frandsen}}, \bibinfo {author} {\bibfnamefont {K.~M.}\ \bibnamefont
  {Taddei}}, \bibinfo {author} {\bibfnamefont {D.~E.}\ \bibnamefont {Bugaris}},
  \bibinfo {author} {\bibfnamefont {R.}~\bibnamefont {Stadel}}, \bibinfo
  {author} {\bibfnamefont {M.}~\bibnamefont {Yi}}, \bibinfo {author}
  {\bibfnamefont {A.}~\bibnamefont {Acharya}}, \bibinfo {author} {\bibfnamefont
  {R.}~\bibnamefont {Osborn}}, \bibinfo {author} {\bibfnamefont
  {S.}~\bibnamefont {Rosenkranz}}, \bibinfo {author} {\bibfnamefont
  {O.}~\bibnamefont {Chmaissem}}, \ and\ \bibinfo {author} {\bibfnamefont
  {R.~J.}\ \bibnamefont {Birgeneau}},\ }\href {\doibase
  10.1103/PhysRevB.98.180505} {\bibfield  {journal} {\bibinfo  {journal} {Phys.
  Rev. B}\ }\textbf {\bibinfo {volume} {98}},\ \bibinfo {pages} {180505(R)}
  (\bibinfo {year} {2018})}\BibitemShut {NoStop}%
\bibitem [{\citenamefont {Frandsen}\ \emph {et~al.}(2017)\citenamefont
  {Frandsen}, \citenamefont {Taddei}, \citenamefont {Yi}, \citenamefont
  {Frano}, \citenamefont {Guguchia}, \citenamefont {Yu}, \citenamefont {Si},
  \citenamefont {Bugaris}, \citenamefont {Stadel}, \citenamefont {Osborn},
  \citenamefont {Rosenkranz}, \citenamefont {Chmaissem},\ and\ \citenamefont
  {Birgeneau}}]{PhysRevLett.119.187001}%
  \BibitemOpen
  \bibfield  {author} {\bibinfo {author} {\bibfnamefont {B.~A.}\ \bibnamefont
  {Frandsen}}, \bibinfo {author} {\bibfnamefont {K.~M.}\ \bibnamefont
  {Taddei}}, \bibinfo {author} {\bibfnamefont {M.}~\bibnamefont {Yi}}, \bibinfo
  {author} {\bibfnamefont {A.}~\bibnamefont {Frano}}, \bibinfo {author}
  {\bibfnamefont {Z.}~\bibnamefont {Guguchia}}, \bibinfo {author}
  {\bibfnamefont {R.}~\bibnamefont {Yu}}, \bibinfo {author} {\bibfnamefont
  {Q.}~\bibnamefont {Si}}, \bibinfo {author} {\bibfnamefont {D.~E.}\
  \bibnamefont {Bugaris}}, \bibinfo {author} {\bibfnamefont {R.}~\bibnamefont
  {Stadel}}, \bibinfo {author} {\bibfnamefont {R.}~\bibnamefont {Osborn}},
  \bibinfo {author} {\bibfnamefont {S.}~\bibnamefont {Rosenkranz}}, \bibinfo
  {author} {\bibfnamefont {O.}~\bibnamefont {Chmaissem}}, \ and\ \bibinfo
  {author} {\bibfnamefont {R.~J.}\ \bibnamefont {Birgeneau}},\ }\href {\doibase
  10.1103/PhysRevLett.119.187001} {\bibfield  {journal} {\bibinfo  {journal}
  {Phys. Rev. Lett.}\ }\textbf {\bibinfo {volume} {119}},\ \bibinfo {pages}
  {187001} (\bibinfo {year} {2017})}\BibitemShut {NoStop}%
\bibitem [{\citenamefont {Okamoto}\ \emph {et~al.}(2010)\citenamefont
  {Okamoto}, \citenamefont {S\'en\'echal}, \citenamefont {Civelli},\ and\
  \citenamefont {Tremblay}}]{PhysRevB.82.180511}%
  \BibitemOpen
  \bibfield  {author} {\bibinfo {author} {\bibfnamefont {S.}~\bibnamefont
  {Okamoto}}, \bibinfo {author} {\bibfnamefont {D.}~\bibnamefont
  {S\'en\'echal}}, \bibinfo {author} {\bibfnamefont {M.}~\bibnamefont
  {Civelli}}, \ and\ \bibinfo {author} {\bibfnamefont {A.-M.~S.}\ \bibnamefont
  {Tremblay}},\ }\href {\doibase 10.1103/PhysRevB.82.180511} {\bibfield
  {journal} {\bibinfo  {journal} {Phys. Rev. B}\ }\textbf {\bibinfo {volume}
  {82}},\ \bibinfo {pages} {180511(R)} (\bibinfo {year} {2010})}\BibitemShut
  {NoStop}%
\bibitem [{\citenamefont {Kushnirenko}\ \emph {et~al.}(2018)\citenamefont
  {Kushnirenko}, \citenamefont {Evtushinsky}, \citenamefont {Kim},
  \citenamefont {Morozov}, \citenamefont {Harnagea}, \citenamefont {Wurmehl},
  \citenamefont {Aswartham}, \citenamefont {Chubukov},\ and\ \citenamefont
  {Borisenko}}]{kushnirenko2018superconductivity}%
  \BibitemOpen
  \bibfield  {author} {\bibinfo {author} {\bibfnamefont {Y.}~\bibnamefont
  {Kushnirenko}}, \bibinfo {author} {\bibfnamefont {D.}~\bibnamefont
  {Evtushinsky}}, \bibinfo {author} {\bibfnamefont {T.}~\bibnamefont {Kim}},
  \bibinfo {author} {\bibfnamefont {I.}~\bibnamefont {Morozov}}, \bibinfo
  {author} {\bibfnamefont {L.}~\bibnamefont {Harnagea}}, \bibinfo {author}
  {\bibfnamefont {S.}~\bibnamefont {Wurmehl}}, \bibinfo {author} {\bibfnamefont
  {S.}~\bibnamefont {Aswartham}}, \bibinfo {author} {\bibfnamefont
  {A.}~\bibnamefont {Chubukov}}, \ and\ \bibinfo {author} {\bibfnamefont
  {S.}~\bibnamefont {Borisenko}},\ }\href@noop {} {\bibfield  {journal}
  {\bibinfo  {journal} {arXiv preprint arXiv:1810.04446}\ } (\bibinfo {year}
  {2018})}\BibitemShut {NoStop}%
\bibitem [{\citenamefont {Koch}\ \emph {et~al.}(2019)\citenamefont {Koch},
  \citenamefont {Konstantinova}, \citenamefont {Abeykoon}, \citenamefont
  {Wang}, \citenamefont {Petrovic}, \citenamefont {Zhu}, \citenamefont
  {Bozin},\ and\ \citenamefont {Billinge}}]{koch2019room}%
  \BibitemOpen
  \bibfield  {author} {\bibinfo {author} {\bibfnamefont {R.~J.}\ \bibnamefont
  {Koch}}, \bibinfo {author} {\bibfnamefont {T.}~\bibnamefont {Konstantinova}},
  \bibinfo {author} {\bibfnamefont {M.}~\bibnamefont {Abeykoon}}, \bibinfo
  {author} {\bibfnamefont {A.}~\bibnamefont {Wang}}, \bibinfo {author}
  {\bibfnamefont {C.}~\bibnamefont {Petrovic}}, \bibinfo {author}
  {\bibfnamefont {Y.}~\bibnamefont {Zhu}}, \bibinfo {author} {\bibfnamefont
  {E.~S.}\ \bibnamefont {Bozin}}, \ and\ \bibinfo {author} {\bibfnamefont
  {S.~J.~L.}\ \bibnamefont {Billinge}},\ }\href {\doibase
  10.1103/PhysRevB.100.020501} {\bibfield  {journal} {\bibinfo  {journal}
  {Phys. Rev. B}\ }\textbf {\bibinfo {volume} {100}},\ \bibinfo {pages}
  {020501(R)} (\bibinfo {year} {2019})}\BibitemShut {NoStop}%
\bibitem [{\citenamefont {Yim}\ \emph {et~al.}(2018)\citenamefont {Yim},
  \citenamefont {Trainer}, \citenamefont {Aluru}, \citenamefont {Chi},
  \citenamefont {Hardy}, \citenamefont {Liang}, \citenamefont {Bonn},\ and\
  \citenamefont {Wahl}}]{Yim2018}%
  \BibitemOpen
  \bibfield  {author} {\bibinfo {author} {\bibfnamefont {C.~M.}\ \bibnamefont
  {Yim}}, \bibinfo {author} {\bibfnamefont {C.}~\bibnamefont {Trainer}},
  \bibinfo {author} {\bibfnamefont {R.}~\bibnamefont {Aluru}}, \bibinfo
  {author} {\bibfnamefont {S.}~\bibnamefont {Chi}}, \bibinfo {author}
  {\bibfnamefont {W.~N.}\ \bibnamefont {Hardy}}, \bibinfo {author}
  {\bibfnamefont {R.}~\bibnamefont {Liang}}, \bibinfo {author} {\bibfnamefont
  {D.}~\bibnamefont {Bonn}}, \ and\ \bibinfo {author} {\bibfnamefont
  {P.}~\bibnamefont {Wahl}},\ }\href@noop {} {\bibfield  {journal} {\bibinfo
  {journal} {Nature Communications}\ }\textbf {\bibinfo {volume} {9}},\
  \bibinfo {pages} {2602} (\bibinfo {year} {2018})}\BibitemShut {NoStop}%
\bibitem [{\citenamefont {Egami}\ and\ \citenamefont
  {Billinge}(2003)}]{Egami2003}%
  \BibitemOpen
  \bibfield  {author} {\bibinfo {author} {\bibfnamefont {T.}~\bibnamefont
  {Egami}}\ and\ \bibinfo {author} {\bibfnamefont {S.~J.~L.}\ \bibnamefont
  {Billinge}},\ }\href@noop {} {\emph {\bibinfo {title} {Underneath the {Bragg}
  Peaks: Structural Analysis of Complex Materials}}}\ (\bibinfo  {publisher}
  {Pergamon Press Elsevier},\ \bibinfo {address} {Oxford, England},\ \bibinfo
  {year} {2003})\BibitemShut {NoStop}%
\bibitem [{\citenamefont {Farrow}\ \emph
  {et~al.}(2007{\natexlab{a}})\citenamefont {Farrow}, \citenamefont {Juhas},
  \citenamefont {Liu}, \citenamefont {Bryndin}, \citenamefont {Bo{\v{z}}in},
  \citenamefont {Bloch}, \citenamefont {Proffen},\ and\ \citenamefont
  {Billinge}}]{Farrow_2007}%
  \BibitemOpen
  \bibfield  {author} {\bibinfo {author} {\bibfnamefont {C.~L.}\ \bibnamefont
  {Farrow}}, \bibinfo {author} {\bibfnamefont {P.}~\bibnamefont {Juhas}},
  \bibinfo {author} {\bibfnamefont {J.~W.}\ \bibnamefont {Liu}}, \bibinfo
  {author} {\bibfnamefont {D.}~\bibnamefont {Bryndin}}, \bibinfo {author}
  {\bibfnamefont {E.~S.}\ \bibnamefont {Bo{\v{z}}in}}, \bibinfo {author}
  {\bibfnamefont {J.}~\bibnamefont {Bloch}}, \bibinfo {author} {\bibfnamefont
  {T.}~\bibnamefont {Proffen}}, \ and\ \bibinfo {author} {\bibfnamefont
  {S.~J.~L.}\ \bibnamefont {Billinge}},\ }\href@noop {} {\bibfield  {journal}
  {\bibinfo  {journal} {Journal of Physics: Condensed Matter}\ }\textbf
  {\bibinfo {volume} {19}},\ \bibinfo {pages} {335219} (\bibinfo {year}
  {2007}{\natexlab{a}})}\BibitemShut {NoStop}%
\bibitem [{\citenamefont {Frandsen}\ \emph {et~al.}(2019)\citenamefont
  {Frandsen}, \citenamefont {Wang}, \citenamefont {Wu}, \citenamefont {Zhao},\
  and\ \citenamefont {Birgeneau}}]{frandsen2019quantitative}%
  \BibitemOpen
  \bibfield  {author} {\bibinfo {author} {\bibfnamefont {B.~A.}\ \bibnamefont
  {Frandsen}}, \bibinfo {author} {\bibfnamefont {Q.}~\bibnamefont {Wang}},
  \bibinfo {author} {\bibfnamefont {S.}~\bibnamefont {Wu}}, \bibinfo {author}
  {\bibfnamefont {J.}~\bibnamefont {Zhao}}, \ and\ \bibinfo {author}
  {\bibfnamefont {R.~J.}\ \bibnamefont {Birgeneau}},\ }\href {\doibase
  10.1103/PhysRevB.100.020504} {\bibfield  {journal} {\bibinfo  {journal}
  {Phys. Rev. B}\ }\textbf {\bibinfo {volume} {100}},\ \bibinfo {pages}
  {020504(R)} (\bibinfo {year} {2019})}\BibitemShut {NoStop}%
\bibitem [{\citenamefont {Louca}\ \emph {et~al.}(2010)\citenamefont {Louca},
  \citenamefont {Horigane}, \citenamefont {Llobet}, \citenamefont {Arita},
  \citenamefont {Ji}, \citenamefont {Katayama}, \citenamefont {Konbu},
  \citenamefont {Nakamura}, \citenamefont {Koo}, \citenamefont {Tong},\ and\
  \citenamefont {Yamada}}]{PhysRevB.81.134524}%
  \BibitemOpen
  \bibfield  {author} {\bibinfo {author} {\bibfnamefont {D.}~\bibnamefont
  {Louca}}, \bibinfo {author} {\bibfnamefont {K.}~\bibnamefont {Horigane}},
  \bibinfo {author} {\bibfnamefont {A.}~\bibnamefont {Llobet}}, \bibinfo
  {author} {\bibfnamefont {R.}~\bibnamefont {Arita}}, \bibinfo {author}
  {\bibfnamefont {S.}~\bibnamefont {Ji}}, \bibinfo {author} {\bibfnamefont
  {N.}~\bibnamefont {Katayama}}, \bibinfo {author} {\bibfnamefont
  {S.}~\bibnamefont {Konbu}}, \bibinfo {author} {\bibfnamefont
  {K.}~\bibnamefont {Nakamura}}, \bibinfo {author} {\bibfnamefont {T.-Y.}\
  \bibnamefont {Koo}}, \bibinfo {author} {\bibfnamefont {P.}~\bibnamefont
  {Tong}}, \ and\ \bibinfo {author} {\bibfnamefont {K.}~\bibnamefont
  {Yamada}},\ }\href {\doibase 10.1103/PhysRevB.81.134524} {\bibfield
  {journal} {\bibinfo  {journal} {Phys. Rev. B}\ }\textbf {\bibinfo {volume}
  {81}},\ \bibinfo {pages} {134524} (\bibinfo {year} {2010})}\BibitemShut
  {NoStop}%
\bibitem [{\citenamefont {Horigane}\ \emph
  {et~al.}(2014{\natexlab{a}})\citenamefont {Horigane}, \citenamefont
  {Kawashima}, \citenamefont {Ji}, \citenamefont {Yoshikawa}, \citenamefont
  {Louca},\ and\ \citenamefont {Akimitsu}}]{horigane2014local}%
  \BibitemOpen
  \bibfield  {author} {\bibinfo {author} {\bibfnamefont {K.}~\bibnamefont
  {Horigane}}, \bibinfo {author} {\bibfnamefont {K.}~\bibnamefont {Kawashima}},
  \bibinfo {author} {\bibfnamefont {S.}~\bibnamefont {Ji}}, \bibinfo {author}
  {\bibfnamefont {M.}~\bibnamefont {Yoshikawa}}, \bibinfo {author}
  {\bibfnamefont {D.}~\bibnamefont {Louca}}, \ and\ \bibinfo {author}
  {\bibfnamefont {J.}~\bibnamefont {Akimitsu}},\ }in\ \href@noop {} {\emph
  {\bibinfo {booktitle} {Proceedings of the International Conference on
  Strongly Correlated Electron Systems (SCES2013)}}}\ (\bibinfo {year} {2014})\
  p.\ \bibinfo {pages} {015039}\BibitemShut {NoStop}%
\bibitem [{\citenamefont {Toby}\ and\ \citenamefont
  {Von~Dreele}(2013)}]{Toby:aj5212}%
  \BibitemOpen
  \bibfield  {author} {\bibinfo {author} {\bibfnamefont {B.~H.}\ \bibnamefont
  {Toby}}\ and\ \bibinfo {author} {\bibfnamefont {R.~B.}\ \bibnamefont
  {Von~Dreele}},\ }\href {\doibase 10.1107/S0021889813003531} {\bibfield
  {journal} {\bibinfo  {journal} {Journal of Applied Crystallography}\ }\textbf
  {\bibinfo {volume} {46}},\ \bibinfo {pages} {544} (\bibinfo {year}
  {2013})}\BibitemShut {NoStop}%
\bibitem [{\citenamefont {Juh{\'{a}}s}\ \emph {et~al.}(2018)\citenamefont
  {Juh{\'{a}}s}, \citenamefont {Louwen}, \citenamefont {van Eijck},
  \citenamefont {Vogt},\ and\ \citenamefont {Billinge}}]{Juhas:po5132}%
  \BibitemOpen
  \bibfield  {author} {\bibinfo {author} {\bibfnamefont {P.}~\bibnamefont
  {Juh{\'{a}}s}}, \bibinfo {author} {\bibfnamefont {J.~N.}\ \bibnamefont
  {Louwen}}, \bibinfo {author} {\bibfnamefont {L.}~\bibnamefont {van Eijck}},
  \bibinfo {author} {\bibfnamefont {E.~T.~C.}\ \bibnamefont {Vogt}}, \ and\
  \bibinfo {author} {\bibfnamefont {S.~J.~L.}\ \bibnamefont {Billinge}},\
  }\href@noop {} {\bibfield  {journal} {\bibinfo  {journal} {Journal of Applied
  Crystallography}\ }\textbf {\bibinfo {volume} {51}},\ \bibinfo {pages} {1492}
  (\bibinfo {year} {2018})}\BibitemShut {NoStop}%
\bibitem [{\citenamefont {Proffen}\ and\ \citenamefont
  {Billinge}(1999)}]{Proffen:gl0603}%
  \BibitemOpen
  \bibfield  {author} {\bibinfo {author} {\bibfnamefont {T.}~\bibnamefont
  {Proffen}}\ and\ \bibinfo {author} {\bibfnamefont {S.~J.~L.}\ \bibnamefont
  {Billinge}},\ }\href {\doibase 10.1107/S0021889899003532} {\bibfield
  {journal} {\bibinfo  {journal} {Journal of Applied Crystallography}\ }\textbf
  {\bibinfo {volume} {32}},\ \bibinfo {pages} {572} (\bibinfo {year}
  {1999})}\BibitemShut {NoStop}%
\bibitem [{\citenamefont {White}\ \emph {et~al.}(2010)\citenamefont {White},
  \citenamefont {Provis}, \citenamefont {Proffen},\ and\ \citenamefont
  {Van~Deventer}}]{white2010effects}%
  \BibitemOpen
  \bibfield  {author} {\bibinfo {author} {\bibfnamefont {C.~E.}\ \bibnamefont
  {White}}, \bibinfo {author} {\bibfnamefont {J.~L.}\ \bibnamefont {Provis}},
  \bibinfo {author} {\bibfnamefont {T.}~\bibnamefont {Proffen}}, \ and\
  \bibinfo {author} {\bibfnamefont {J.~S.}\ \bibnamefont {Van~Deventer}},\
  }\href@noop {} {\bibfield  {journal} {\bibinfo  {journal} {Journal of the
  American Ceramic Society}\ }\textbf {\bibinfo {volume} {93}},\ \bibinfo
  {pages} {3486} (\bibinfo {year} {2010})}\BibitemShut {NoStop}%
\bibitem [{\citenamefont {Pa{\'s}ciak}\ \emph {et~al.}(2019)\citenamefont
  {Pa{\'s}ciak}, \citenamefont {Ondrejkovic}, \citenamefont {Kulda},
  \citenamefont {Van{\v{e}}k}, \citenamefont {Drahokoupil}, \citenamefont
  {Steciuk}, \citenamefont {Palatinus}, \citenamefont {Welberry}, \citenamefont
  {Fischer}, \citenamefont {Hlinka},\ and\ \citenamefont
  {Buixaderas}}]{pasciak2019local}%
  \BibitemOpen
  \bibfield  {author} {\bibinfo {author} {\bibfnamefont {M.}~\bibnamefont
  {Pa{\'s}ciak}}, \bibinfo {author} {\bibfnamefont {P.}~\bibnamefont
  {Ondrejkovic}}, \bibinfo {author} {\bibfnamefont {J.}~\bibnamefont {Kulda}},
  \bibinfo {author} {\bibfnamefont {P.}~\bibnamefont {Van{\v{e}}k}}, \bibinfo
  {author} {\bibfnamefont {J.}~\bibnamefont {Drahokoupil}}, \bibinfo {author}
  {\bibfnamefont {G.}~\bibnamefont {Steciuk}}, \bibinfo {author} {\bibfnamefont
  {L.}~\bibnamefont {Palatinus}}, \bibinfo {author} {\bibfnamefont {T.~R.}\
  \bibnamefont {Welberry}}, \bibinfo {author} {\bibfnamefont {H.~E.}\
  \bibnamefont {Fischer}}, \bibinfo {author} {\bibfnamefont {J.}~\bibnamefont
  {Hlinka}}, \ and\ \bibinfo {author} {\bibfnamefont {E.}~\bibnamefont
  {Buixaderas}},\ }\href@noop {} {\bibfield  {journal} {\bibinfo  {journal}
  {Physical Review B}\ }\textbf {\bibinfo {volume} {99}},\ \bibinfo {pages}
  {104102} (\bibinfo {year} {2019})}\BibitemShut {NoStop}%
\bibitem [{\citenamefont {Slawinski}\ \emph {et~al.}(2019)\citenamefont
  {Slawinski}, \citenamefont {Playford}, \citenamefont {Hull}, \citenamefont
  {Norberg}, \citenamefont {Eriksson}, \citenamefont {Gustafsson},
  \citenamefont {Edstrom},\ and\ \citenamefont {Brant}}]{slawinski2019neutron}%
  \BibitemOpen
  \bibfield  {author} {\bibinfo {author} {\bibfnamefont {W.~A.}\ \bibnamefont
  {Slawinski}}, \bibinfo {author} {\bibfnamefont {H.~Y.}\ \bibnamefont
  {Playford}}, \bibinfo {author} {\bibfnamefont {S.}~\bibnamefont {Hull}},
  \bibinfo {author} {\bibfnamefont {S.~T.}\ \bibnamefont {Norberg}}, \bibinfo
  {author} {\bibfnamefont {S.~G.}\ \bibnamefont {Eriksson}}, \bibinfo {author}
  {\bibfnamefont {T.}~\bibnamefont {Gustafsson}}, \bibinfo {author}
  {\bibfnamefont {K.}~\bibnamefont {Edstrom}}, \ and\ \bibinfo {author}
  {\bibfnamefont {W.~R.}\ \bibnamefont {Brant}},\ }\href@noop {} {\bibfield
  {journal} {\bibinfo  {journal} {Chemistry of Materials}\ }\textbf {\bibinfo
  {volume} {31}},\ \bibinfo {pages} {5024} (\bibinfo {year}
  {2019})}\BibitemShut {NoStop}%
\bibitem [{\citenamefont {Jeong}\ and\ \citenamefont
  {Kim}(2019)}]{jeong2019high}%
  \BibitemOpen
  \bibfield  {author} {\bibinfo {author} {\bibfnamefont {I.-K.}\ \bibnamefont
  {Jeong}}\ and\ \bibinfo {author} {\bibfnamefont {B.-G.}\ \bibnamefont
  {Kim}},\ }\href@noop {} {\bibfield  {journal} {\bibinfo  {journal} {Journal
  of Applied Physics}\ }\textbf {\bibinfo {volume} {126}},\ \bibinfo {pages}
  {014101} (\bibinfo {year} {2019})}\BibitemShut {NoStop}%
\bibitem [{\citenamefont {Masadeh}\ \emph {et~al.}(2007)\citenamefont
  {Masadeh}, \citenamefont {Bo\ifmmode~\check{z}\else \v{z}\fi{}in},
  \citenamefont {Farrow}, \citenamefont {Paglia}, \citenamefont {Juhas},
  \citenamefont {Billinge}, \citenamefont {Karkamkar},\ and\ \citenamefont
  {Kanatzidis}}]{masadeh2007quantitative}%
  \BibitemOpen
  \bibfield  {author} {\bibinfo {author} {\bibfnamefont {A.~S.}\ \bibnamefont
  {Masadeh}}, \bibinfo {author} {\bibfnamefont {E.~S.}\ \bibnamefont
  {Bo\ifmmode~\check{z}\else \v{z}\fi{}in}}, \bibinfo {author} {\bibfnamefont
  {C.~L.}\ \bibnamefont {Farrow}}, \bibinfo {author} {\bibfnamefont
  {G.}~\bibnamefont {Paglia}}, \bibinfo {author} {\bibfnamefont
  {P.}~\bibnamefont {Juhas}}, \bibinfo {author} {\bibfnamefont {S.~J.~L.}\
  \bibnamefont {Billinge}}, \bibinfo {author} {\bibfnamefont {A.}~\bibnamefont
  {Karkamkar}}, \ and\ \bibinfo {author} {\bibfnamefont {M.~G.}\ \bibnamefont
  {Kanatzidis}},\ }\href@noop {} {\bibfield  {journal} {\bibinfo  {journal}
  {Physical Review B}\ }\textbf {\bibinfo {volume} {76}},\ \bibinfo {pages}
  {115413} (\bibinfo {year} {2007})}\BibitemShut {NoStop}%
\bibitem [{\citenamefont {Toby}\ and\ \citenamefont
  {Egami}(1992)}]{toby1992accuracy}%
  \BibitemOpen
  \bibfield  {author} {\bibinfo {author} {\bibfnamefont {B.}~\bibnamefont
  {Toby}}\ and\ \bibinfo {author} {\bibfnamefont {T.}~\bibnamefont {Egami}},\
  }\href@noop {} {\bibfield  {journal} {\bibinfo  {journal} {Acta
  Crystallographica Section A: Foundations of Crystallography}\ }\textbf
  {\bibinfo {volume} {48}},\ \bibinfo {pages} {336} (\bibinfo {year}
  {1992})}\BibitemShut {NoStop}%
\bibitem [{\citenamefont {Kr\"{o}ncke}\ \emph {et~al.}(2008)\citenamefont
  {Kr\"{o}ncke}, \citenamefont {Figge}, \citenamefont {Hommel},\ and\
  \citenamefont {Epelbaum}}]{einsteinplot}%
  \BibitemOpen
  \bibfield  {author} {\bibinfo {author} {\bibfnamefont {H.}~\bibnamefont
  {Kr\"{o}ncke}}, \bibinfo {author} {\bibfnamefont {S.}~\bibnamefont {Figge}},
  \bibinfo {author} {\bibfnamefont {D.}~\bibnamefont {Hommel}}, \ and\ \bibinfo
  {author} {\bibfnamefont {B.~M.}\ \bibnamefont {Epelbaum}},\ }\href@noop {}
  {\bibfield  {journal} {\bibinfo  {journal} {Acta Phys. Pol. A}\ }\textbf
  {\bibinfo {volume} {114}},\ \bibinfo {pages} {1193} (\bibinfo {year}
  {2008})}\BibitemShut {NoStop}%
\bibitem [{\citenamefont {Farrow}\ \emph
  {et~al.}(2007{\natexlab{b}})\citenamefont {Farrow}, \citenamefont {Juhas},
  \citenamefont {Liu}, \citenamefont {Bryndin}, \citenamefont {Bo{\v{z}}in},
  \citenamefont {Bloch}, \citenamefont {Proffen},\ and\ \citenamefont
  {Billinge}}]{farrow2007pdffit2}%
  \BibitemOpen
  \bibfield  {author} {\bibinfo {author} {\bibfnamefont {C.}~\bibnamefont
  {Farrow}}, \bibinfo {author} {\bibfnamefont {P.}~\bibnamefont {Juhas}},
  \bibinfo {author} {\bibfnamefont {J.}~\bibnamefont {Liu}}, \bibinfo {author}
  {\bibfnamefont {D.}~\bibnamefont {Bryndin}}, \bibinfo {author} {\bibfnamefont
  {E.}~\bibnamefont {Bo{\v{z}}in}}, \bibinfo {author} {\bibfnamefont
  {J.}~\bibnamefont {Bloch}}, \bibinfo {author} {\bibfnamefont
  {T.}~\bibnamefont {Proffen}}, \ and\ \bibinfo {author} {\bibfnamefont
  {S.}~\bibnamefont {Billinge}},\ }\href@noop {} {\bibfield  {journal}
  {\bibinfo  {journal} {Journal of Physics: Condensed Matter}\ }\textbf
  {\bibinfo {volume} {19}},\ \bibinfo {pages} {335219} (\bibinfo {year}
  {2007}{\natexlab{b}})}\BibitemShut {NoStop}%
\bibitem [{\citenamefont {Abeykoon}\ \emph {et~al.}(2009)\citenamefont
  {Abeykoon}, \citenamefont {Donner}, \citenamefont {Brunelli}, \citenamefont
  {Castro-Colin}, \citenamefont {Jacobson},\ and\ \citenamefont
  {Moss}}]{abeykoon2009average}%
  \BibitemOpen
  \bibfield  {author} {\bibinfo {author} {\bibfnamefont {A.~M.}\ \bibnamefont
  {Abeykoon}}, \bibinfo {author} {\bibfnamefont {W.}~\bibnamefont {Donner}},
  \bibinfo {author} {\bibfnamefont {M.}~\bibnamefont {Brunelli}}, \bibinfo
  {author} {\bibfnamefont {M.}~\bibnamefont {Castro-Colin}}, \bibinfo {author}
  {\bibfnamefont {A.~J.}\ \bibnamefont {Jacobson}}, \ and\ \bibinfo {author}
  {\bibfnamefont {S.~C.}\ \bibnamefont {Moss}},\ }\href@noop {} {\bibfield
  {journal} {\bibinfo  {journal} {Journal of the American Chemical Society}\
  }\textbf {\bibinfo {volume} {131}},\ \bibinfo {pages} {13230} (\bibinfo
  {year} {2009})}\BibitemShut {NoStop}%
\bibitem [{\citenamefont {{OriginPro 2013}}()}]{origin}%
  \BibitemOpen
  \bibfield  {author} {\bibinfo {author} {\bibnamefont {{OriginPro 2013}}},\
  }\href {www. originlab. com.} {\emph {\bibinfo {title} {{OriginLab
  Corporation}}}},\ \bibinfo {address} {Northampton, MA, USA}\BibitemShut
  {NoStop}%
\bibitem [{\citenamefont {McCabe}\ \emph
  {et~al.}(2014{\natexlab{b}})\citenamefont {McCabe}, \citenamefont {Wills},
  \citenamefont {Chapon}, \citenamefont {Manuel},\ and\ \citenamefont
  {Evans}}]{PhysRevB.90.165111}%
  \BibitemOpen
  \bibfield  {author} {\bibinfo {author} {\bibfnamefont {E.~E.}\ \bibnamefont
  {McCabe}}, \bibinfo {author} {\bibfnamefont {A.~S.}\ \bibnamefont {Wills}},
  \bibinfo {author} {\bibfnamefont {L.}~\bibnamefont {Chapon}}, \bibinfo
  {author} {\bibfnamefont {P.}~\bibnamefont {Manuel}}, \ and\ \bibinfo {author}
  {\bibfnamefont {J.~S.~O.}\ \bibnamefont {Evans}},\ }\href {\doibase
  10.1103/PhysRevB.90.165111} {\bibfield  {journal} {\bibinfo  {journal} {Phys.
  Rev. B}\ }\textbf {\bibinfo {volume} {90}},\ \bibinfo {pages} {165111}
  (\bibinfo {year} {2014}{\natexlab{b}})}\BibitemShut {NoStop}%
\bibitem [{\citenamefont {Oogarah}\ \emph {et~al.}(2017)\citenamefont
  {Oogarah}, \citenamefont {Stockdale}, \citenamefont {Stock}, \citenamefont
  {Evans}, \citenamefont {Wills}, \citenamefont {Taylor},\ and\ \citenamefont
  {McCabe}}]{PhysRevB.95.174441}%
  \BibitemOpen
  \bibfield  {author} {\bibinfo {author} {\bibfnamefont {R.~K.}\ \bibnamefont
  {Oogarah}}, \bibinfo {author} {\bibfnamefont {C.~P.~J.}\ \bibnamefont
  {Stockdale}}, \bibinfo {author} {\bibfnamefont {C.}~\bibnamefont {Stock}},
  \bibinfo {author} {\bibfnamefont {J.~S.~O.}\ \bibnamefont {Evans}}, \bibinfo
  {author} {\bibfnamefont {A.~S.}\ \bibnamefont {Wills}}, \bibinfo {author}
  {\bibfnamefont {J.~W.}\ \bibnamefont {Taylor}}, \ and\ \bibinfo {author}
  {\bibfnamefont {E.~E.}\ \bibnamefont {McCabe}},\ }\href {\doibase
  10.1103/PhysRevB.95.174441} {\bibfield  {journal} {\bibinfo  {journal} {Phys.
  Rev. B}\ }\textbf {\bibinfo {volume} {95}},\ \bibinfo {pages} {174441}
  (\bibinfo {year} {2017})}\BibitemShut {NoStop}%
\bibitem [{\citenamefont {Horigane}\ \emph
  {et~al.}(2014{\natexlab{b}})\citenamefont {Horigane}, \citenamefont
  {Llobet},\ and\ \citenamefont {Louca}}]{PhysRevLett.112.097001}%
  \BibitemOpen
  \bibfield  {author} {\bibinfo {author} {\bibfnamefont {K.}~\bibnamefont
  {Horigane}}, \bibinfo {author} {\bibfnamefont {A.}~\bibnamefont {Llobet}}, \
  and\ \bibinfo {author} {\bibfnamefont {D.}~\bibnamefont {Louca}},\ }\href
  {\doibase 10.1103/PhysRevLett.112.097001} {\bibfield  {journal} {\bibinfo
  {journal} {Phys. Rev. Lett.}\ }\textbf {\bibinfo {volume} {112}},\ \bibinfo
  {pages} {097001} (\bibinfo {year} {2014}{\natexlab{b}})}\BibitemShut
  {NoStop}%
\bibitem [{\citenamefont {Yi}\ \emph {et~al.}(2019)\citenamefont {Yi},
  \citenamefont {Pfau}, \citenamefont {Zhang}, \citenamefont {He},
  \citenamefont {Wu}, \citenamefont {Chen}, \citenamefont {Ye}, \citenamefont
  {Hashimoto}, \citenamefont {Yu}, \citenamefont {Si}, \citenamefont {Lee},
  \citenamefont {Dai}, \citenamefont {Shen}, \citenamefont {Lu},\ and\
  \citenamefont {Birgeneau}}]{PhysRevX.9.041049}%
  \BibitemOpen
  \bibfield  {author} {\bibinfo {author} {\bibfnamefont {M.}~\bibnamefont
  {Yi}}, \bibinfo {author} {\bibfnamefont {H.}~\bibnamefont {Pfau}}, \bibinfo
  {author} {\bibfnamefont {Y.}~\bibnamefont {Zhang}}, \bibinfo {author}
  {\bibfnamefont {Y.}~\bibnamefont {He}}, \bibinfo {author} {\bibfnamefont
  {H.}~\bibnamefont {Wu}}, \bibinfo {author} {\bibfnamefont {T.}~\bibnamefont
  {Chen}}, \bibinfo {author} {\bibfnamefont {Z.~R.}\ \bibnamefont {Ye}},
  \bibinfo {author} {\bibfnamefont {M.}~\bibnamefont {Hashimoto}}, \bibinfo
  {author} {\bibfnamefont {R.}~\bibnamefont {Yu}}, \bibinfo {author}
  {\bibfnamefont {Q.}~\bibnamefont {Si}}, \bibinfo {author} {\bibfnamefont
  {D.-H.}\ \bibnamefont {Lee}}, \bibinfo {author} {\bibfnamefont
  {P.}~\bibnamefont {Dai}}, \bibinfo {author} {\bibfnamefont {Z.-X.}\
  \bibnamefont {Shen}}, \bibinfo {author} {\bibfnamefont {D.~H.}\ \bibnamefont
  {Lu}}, \ and\ \bibinfo {author} {\bibfnamefont {R.~J.}\ \bibnamefont
  {Birgeneau}},\ }\href {\doibase 10.1103/PhysRevX.9.041049} {\bibfield
  {journal} {\bibinfo  {journal} {Phys. Rev. X}\ }\textbf {\bibinfo {volume}
  {9}},\ \bibinfo {pages} {041049} (\bibinfo {year} {2019})}\BibitemShut
  {NoStop}%
\bibitem [{\citenamefont {Kuo}\ \emph {et~al.}(2016)\citenamefont {Kuo},
  \citenamefont {Chu}, \citenamefont {Palmstrom}, \citenamefont {Kivelson},\
  and\ \citenamefont {Fisher}}]{Kuo958}%
  \BibitemOpen
  \bibfield  {author} {\bibinfo {author} {\bibfnamefont {H.-H.}\ \bibnamefont
  {Kuo}}, \bibinfo {author} {\bibfnamefont {J.-H.}\ \bibnamefont {Chu}},
  \bibinfo {author} {\bibfnamefont {J.~C.}\ \bibnamefont {Palmstrom}}, \bibinfo
  {author} {\bibfnamefont {S.~A.}\ \bibnamefont {Kivelson}}, \ and\ \bibinfo
  {author} {\bibfnamefont {I.~R.}\ \bibnamefont {Fisher}},\ }\href {\doibase
  10.1126/science.aab0103} {\bibfield  {journal} {\bibinfo  {journal}
  {Science}\ }\textbf {\bibinfo {volume} {352}},\ \bibinfo {pages} {958}
  (\bibinfo {year} {2016})}\BibitemShut {NoStop}%
\bibitem [{\citenamefont {Böhmer}\ and\ \citenamefont
  {Meingast}(2016)}]{BOHMER201690}%
  \BibitemOpen
  \bibfield  {author} {\bibinfo {author} {\bibfnamefont {A.~E.}\ \bibnamefont
  {Böhmer}}\ and\ \bibinfo {author} {\bibfnamefont {C.}~\bibnamefont
  {Meingast}},\ }\href {\doibase https://doi.org/10.1016/j.crhy.2015.07.001}
  {\bibfield  {journal} {\bibinfo  {journal} {Comptes Rendus Physique}\
  }\textbf {\bibinfo {volume} {17}},\ \bibinfo {pages} {90 } (\bibinfo {year}
  {2016})}\BibitemShut {NoStop}%
\bibitem [{\citenamefont {Rosenthal}\ \emph {et~al.}(2014)\citenamefont
  {Rosenthal}, \citenamefont {Andrade}, \citenamefont {Arguello}, \citenamefont
  {Fernandes}, \citenamefont {Xing}, \citenamefont {Wang}, \citenamefont {Jin},
  \citenamefont {Millis},\ and\ \citenamefont {Pasupathy}}]{Rosenthal2014}%
  \BibitemOpen
  \bibfield  {author} {\bibinfo {author} {\bibfnamefont {E.~P.}\ \bibnamefont
  {Rosenthal}}, \bibinfo {author} {\bibfnamefont {E.~F.}\ \bibnamefont
  {Andrade}}, \bibinfo {author} {\bibfnamefont {C.~J.}\ \bibnamefont
  {Arguello}}, \bibinfo {author} {\bibfnamefont {R.~M.}\ \bibnamefont
  {Fernandes}}, \bibinfo {author} {\bibfnamefont {L.~Y.}\ \bibnamefont {Xing}},
  \bibinfo {author} {\bibfnamefont {X.~C.}\ \bibnamefont {Wang}}, \bibinfo
  {author} {\bibfnamefont {C.~Q.}\ \bibnamefont {Jin}}, \bibinfo {author}
  {\bibfnamefont {A.~J.}\ \bibnamefont {Millis}}, \ and\ \bibinfo {author}
  {\bibfnamefont {A.~N.}\ \bibnamefont {Pasupathy}},\ }\href
  {https://doi.org/10.1038/nphys2870} {\bibfield  {journal} {\bibinfo
  {journal} {Nature Physics}\ }\textbf {\bibinfo {volume} {10}},\ \bibinfo
  {pages} {225} (\bibinfo {year} {2014})}\BibitemShut {NoStop}%
\bibitem [{\citenamefont {Wang}\ \emph {et~al.}(2017)\citenamefont {Wang},
  \citenamefont {Liu}, \citenamefont {Efremov},\ and\ \citenamefont {van~den
  Brink}}]{PhysRevB.95.024511}%
  \BibitemOpen
  \bibfield  {author} {\bibinfo {author} {\bibfnamefont {J.}~\bibnamefont
  {Wang}}, \bibinfo {author} {\bibfnamefont {G.-Z.}\ \bibnamefont {Liu}},
  \bibinfo {author} {\bibfnamefont {D.~V.}\ \bibnamefont {Efremov}}, \ and\
  \bibinfo {author} {\bibfnamefont {J.}~\bibnamefont {van~den Brink}},\ }\href
  {\doibase 10.1103/PhysRevB.95.024511} {\bibfield  {journal} {\bibinfo
  {journal} {Phys. Rev. B}\ }\textbf {\bibinfo {volume} {95}},\ \bibinfo
  {pages} {024511} (\bibinfo {year} {2017})}\BibitemShut {NoStop}%
\bibitem [{\citenamefont {Dai}(2015)}]{RevModPhys.87.855}%
  \BibitemOpen
  \bibfield  {author} {\bibinfo {author} {\bibfnamefont {P.}~\bibnamefont
  {Dai}},\ }\href {\doibase 10.1103/RevModPhys.87.855} {\bibfield  {journal}
  {\bibinfo  {journal} {Rev. Mod. Phys.}\ }\textbf {\bibinfo {volume} {87}},\
  \bibinfo {pages} {855} (\bibinfo {year} {2015})}\BibitemShut {NoStop}%
\bibitem [{\citenamefont {Stewart}(2011)}]{RevModPhys.83.1589}%
  \BibitemOpen
  \bibfield  {author} {\bibinfo {author} {\bibfnamefont {G.~R.}\ \bibnamefont
  {Stewart}},\ }\href {\doibase 10.1103/RevModPhys.83.1589} {\bibfield
  {journal} {\bibinfo  {journal} {Rev. Mod. Phys.}\ }\textbf {\bibinfo {volume}
  {83}},\ \bibinfo {pages} {1589} (\bibinfo {year} {2011})}\BibitemShut
  {NoStop}%
\bibitem [{\citenamefont {Freelon}\ \emph {et~al.}(2021)\citenamefont
  {Freelon}, \citenamefont {Sarkar}, \citenamefont {Kamusella}, \citenamefont
  {Br{\"u}ckner}, \citenamefont {Grinenko}, \citenamefont {Acharya},
  \citenamefont {Laad}, \citenamefont {Craco}, \citenamefont {Yamani},
  \citenamefont {Flacau}, \citenamefont {Swainson}, \citenamefont {Frandsen},
  \citenamefont {Birgeneau}, \citenamefont {Liu}, \citenamefont {Karki},
  \citenamefont {Alfailakawi}, \citenamefont {Neuefeind}, \citenamefont
  {Everett}, \citenamefont {Wang}, \citenamefont {Xu}, \citenamefont {Fang},\
  and\ \citenamefont {Klauss}}]{Freelon2021}%
  \BibitemOpen
  \bibfield  {author} {\bibinfo {author} {\bibfnamefont {B.}~\bibnamefont
  {Freelon}}, \bibinfo {author} {\bibfnamefont {R.}~\bibnamefont {Sarkar}},
  \bibinfo {author} {\bibfnamefont {S.}~\bibnamefont {Kamusella}}, \bibinfo
  {author} {\bibfnamefont {F.}~\bibnamefont {Br{\"u}ckner}}, \bibinfo {author}
  {\bibfnamefont {V.}~\bibnamefont {Grinenko}}, \bibinfo {author}
  {\bibfnamefont {S.}~\bibnamefont {Acharya}}, \bibinfo {author} {\bibfnamefont
  {M.}~\bibnamefont {Laad}}, \bibinfo {author} {\bibfnamefont {L.}~\bibnamefont
  {Craco}}, \bibinfo {author} {\bibfnamefont {Z.}~\bibnamefont {Yamani}},
  \bibinfo {author} {\bibfnamefont {R.}~\bibnamefont {Flacau}}, \bibinfo
  {author} {\bibfnamefont {I.}~\bibnamefont {Swainson}}, \bibinfo {author}
  {\bibfnamefont {B.}~\bibnamefont {Frandsen}}, \bibinfo {author}
  {\bibfnamefont {R.}~\bibnamefont {Birgeneau}}, \bibinfo {author}
  {\bibfnamefont {Y.}~\bibnamefont {Liu}}, \bibinfo {author} {\bibfnamefont
  {B.}~\bibnamefont {Karki}}, \bibinfo {author} {\bibfnamefont
  {A.}~\bibnamefont {Alfailakawi}}, \bibinfo {author} {\bibfnamefont {J.~C.}\
  \bibnamefont {Neuefeind}}, \bibinfo {author} {\bibfnamefont {M.}~\bibnamefont
  {Everett}}, \bibinfo {author} {\bibfnamefont {H.}~\bibnamefont {Wang}},
  \bibinfo {author} {\bibfnamefont {B.}~\bibnamefont {Xu}}, \bibinfo {author}
  {\bibfnamefont {M.}~\bibnamefont {Fang}}, \ and\ \bibinfo {author}
  {\bibfnamefont {H.-H.}\ \bibnamefont {Klauss}},\ }\href@noop {} {\bibfield
  {journal} {\bibinfo  {journal} {npj Quantum Materials}\ }\textbf {\bibinfo
  {volume} {6}},\ \bibinfo {pages} {4} (\bibinfo {year} {2021})}\BibitemShut
  {NoStop}%
\end{thebibliography}%

\end{document}